\documentclass[twocolumn,showpacs,preprintnumbers,amsmath,amssymb,showkeys]{revtex4}
 \usepackage{graphicx}
 \usepackage{dcolumn}
 \usepackage{bm}

\begin{document}
 \title{ Optical properties of the Q1D multiband models --
 the transverse equation of motion approach}
\author{Ivan Kup\v{c}i\'{c}\footnote{
E-mail address: kupcic@phy.hr
(I. Kup\v{c}i\'{c})}
and  Slaven Bari\v{s}i\'{c}} 
\affiliation{ Department of Physics, 
        Faculty of Science, POB 331,  HR-10 002 Zagreb, Croatia}
%\date{}

\begin{abstract}
The electrodynamic features of the multiband model
are examined using the transverse equation of motion approach 
in order to give the explanation of several long-standing problems.
	It turns out that the exact summation of the most singular terms in
powers of $1/\omega^{n}$ leads to the total optical conductivity
which, in the zero-frequency limit, reduces to the results of the Boltzmann
equation, for both the metallic and semiconducting two-band regime.
	The detailed calculations are carried out for the 
quasi-one-dimensional (Q1D) two-band model corresponding to imperfect
charge-density-wave (CDW) nesting. 
	It is also shown that the present treatment of the impurity 
scattering processes  gives the DC conductivity of the ordered CDW 
state in agreement with the experimental observation.
	Finally,  the DC and optical conductivity are calculated numerically 
for a few typical Q1D cases.
\end{abstract}


\maketitle

 \section{Introduction }
%
%	Sec. 1.
%
Current investigations of the strongly correlated electron systems
often deal with collective contributions to the electrical
conductivity (and to the other transport coefficients).
	This includes in particular the conductivity in the 
charge-density-wave (CDW) or spin-density-wave (SDW) ordered 
states.
	In most of those studies the electrical conductivity is
determined, theoretically and experimentally, relative either
to the conductivity in the metallic state or with respect to the 
conductivity in the semiconducting state of the pinned CDW/SDW.
	Obviously, the prerequisite to such  procedures is the accurate
evaluation of the conductivity in the metallic state or in the 
state of the pinned CDW/SDW, and this simple, yet not completely solved, 
problem is in the focus of our interest here.

Actually, using the microscopic transverse response theory, 
Lee, Rice and Anderson
have found that the single-particle optical conductivity in the ordered
CDW state with the negligible small number 
of scattering centers is given in terms of the semiconducting current-current
correlation function which describes excitations across the gap 
\cite{LRA,Wooten,Gruner}.
	However, it is also  shown that for the typical 
value of the CDW gap and of the zero-frequency damping energy 
(arising from the impurity scattering processes) their result
matches up neither the result of the longitudinal response theory 
\cite{KupcicPB1} 
nor the experimental observation
\cite{Degiorgi,Kim}.
	Yet, the longitudinal and transverse response have to coincide
for fast enough (quasi)homogeneous longitudinal fields, as implicit for
example in the Maxwell equations of the medium, which employ 
only one dielectric function.

In his textbook~\cite{Mahan}, 
Mahan has further shown that an alternative, so-called  force-force
correlation function method gives a good description
of the (high-frequency) optical processes in both metallic and 
semiconducting systems, including various excitations 
within the conduction band and across the (pseudo)gaps, 
but also fails to reach correctly
the $\omega \rightarrow 0$ limit, requiring a specific 
$\omega = 0$ field-theory  approach 
\cite{Mahan,Abrikosov,Doniach}.

Also important is the observation that  most of the transport-coefficient
analyses are based on the Boltzmann equations applied to the nearly-free
electron models
\cite{Mahan,Ziman}, 
completely neglecting band periodicity in the reciprocal space.

Most of these issues can be settled down using the
longitudinal response theory with a particular care devoted
to the continuity equation
\cite{Pines,KupcicUP}.
	In the present article, it will be shown  
that this can be done  alternatively (but  in a somewhat less strict way) 
using the gauge-invariant form of the transverse approach.
	For this purpose, we consider  a  quasi-one-dimensional (Q1D) 
two-band model with the impurity scattering taken into consideration.

The two bands are taken to result from the site-energy dimerization in the
highly conducting direction, and, consequently, the Bloch functions 
and all relevant vertex functions can be determined analytically
(Sec.~III).
	Using the  equation of motion approach (which is found to be the 
generalization of the force-force correlation function approach),
the intra- and interband optical conductivity 
are calculated (Sec.~IV).
	In the intraband channel of  the  transverse 
correlation functions, the most singular processes in powers of
$1/\omega^{n}$ are collected, resulting in the optical conductivity which
matches up the DC conductivity obtained by the Boltzmann equations \cite{Ziman}
or by the Landau response theory \cite{Pines}.
	The resulting interband conductivity in the CDW ordered state is
found to be consistent with the experimental observation. 
	Theorywise, the semiconducting current-current correlation 
function \cite{LRA,Wooten,Gruner}  
is replaced by a slightly modified  function containing 
an additional factor which comes from the gauge invariant treatment 
of the diamagnetic current  contributions.
	Finally (Sec.~V), the optical and DC conductivity  
are determined for a few typical Q1D cases.

 \section{Transverse multiband response theory }
%
%	Sec. 2.
%

The optical conductivity tensor $\sigma_{\alpha \alpha} (\omega)$
is a measure of the absorption rate for  the (transverse) electromagnetic 
waves traveling across the crystal, and
the measured spectra, together with the DC conductivity data
and other transport coefficients, 
are an extremely valuable source of the information 
about the electronic subsystem.
	Although some aspects of the microscopic response theory can be found
in the textbooks
\cite{Wooten,Gruner,Mahan,Abrikosov,Doniach,Pines,Kittel}, 
there is no systematic microscopic solution to the
multiband optical conductivity  problem.
	Actually, it is easy to determine macroscopic symmetry 
features of $\sigma_{\alpha \alpha} (\omega)$, even in a general case.
	In this respect, we shall combine the macroscopic symmetry 
features with the microscopic description of the electron-photon coupling 
functions (determined for a simple, exactly solvable Q1D electronic model)
to develop a consistent microscopic multiband response theory.

 \subsection{Optical conductivity tensor}
%
%	Sec. 2.1
%

The optical conductivity analysis starts with the Hamiltonian
\cite{Pines} 
\begin{eqnarray}
H &=& H^{\rm field}_0 + H^{\rm el}_0 + H^{\rm ext}_1+ H^{\rm ext}_2, 
\label{eq1}
\end{eqnarray}
%
%  (1)
%
which comprises the bare photon term $H^{\rm field}_0$, the bare electronic 
Hamiltonian $H^{\rm el}_0$ and the first-order and the second-order 
electron-photon coupling term, $H^{\rm ext}_1$ and  $H^{\rm ext}_2$.
	The bare photon contribution is
\begin{eqnarray}
H^{\rm field}_0  &=&  \frac{1}{2} \sum_{{\bf q} \alpha}
\big[ P_{{\bf q} \alpha}^{ \dagger} P_{{\bf q} \alpha}
+ \omega_{{\bf q}0}^2 Q_{{\bf q} \alpha}^{ \dagger} Q_{{\bf q} \alpha} 
\big].
\label{eq2}
\end{eqnarray}
%
%  (2)
%
	Here ${\bf q}$ and $\alpha$ are the wave vector and  the polarization 
of the photon field $ Q_{{\bf q} \alpha}$, 
$P_{{\bf q} \alpha}$ is the field conjugate to $ Q_{{\bf q} \alpha}$, 
and $\omega_{{\bf q}0}  = c q$ is the bare 
photon dispersion.
	The structure of a typical Q1D tight-binding electronic 
Hamiltonian $H^{\rm el} = H^{\rm el}_0 + H^{\rm ext}_1+ H^{\rm ext}_2 $
is determined below. 
	However, notice that general symmetry properties of 
$\sigma_{\alpha \alpha} (\omega)$ discussed here do not depend on
details of $H^{\rm el}$.

To obtain $\sigma_{\alpha \alpha} (\omega)$, 
the retarded photon Green function is 
required \cite{footnote1}
\begin{eqnarray}
\langle \langle Q_{{\bf q} \alpha} ; Q_{-{\bf q} \alpha} 
\rangle \rangle_{t} &=&   -{\rm i} 
\Theta(t) \langle \big[ Q_{{\bf q} \alpha}(t), Q_{-{\bf q}} (0)] \rangle
\\ \nonumber 
&=& {\rm e}^{ -\eta t} 
\frac{1}{2 \pi} \int_{-\infty}^{\infty} {\rm d} \omega \,
{\rm e}^{{\rm i} \omega t} 
\langle \langle Q_{{\bf q} \alpha} ; Q_{-{\bf q} \alpha} 
\rangle \rangle_{\omega} ,
\label{eq3}
\end{eqnarray}
%
%  (3)
%
with $ Q_{{\bf q} \alpha}(t) = \
\displaystyle {\rm e}^{{\rm i} H t/\hbar} 
 Q_{{\bf q} \alpha}  {\rm e}^{-{\rm i} H t/\hbar}$
and $ Q_{{\bf q} \alpha}(0) = Q_{{\bf q} \alpha}$ 
representing the photon fields in the Heisenberg picture at 
time $t$ and $t=0$, respectively.
	Using the equation of motion formalism, we get
\begin{eqnarray}
\big[ \omega\big(\omega + {\rm i} \eta \big) - \omega_{{\bf q} \alpha} ^2 \big]
\langle \langle Q_{{\bf q} \alpha} ; Q_{-{\bf q} \alpha} 
\rangle \rangle_{\omega}
 &=& \hbar,
\label{eq4}
\end{eqnarray}
%
%  (4)
%	
with the adiabatic term $\eta \rightarrow 0^+$.
	The renormalized photon frequency $\omega_{{\bf q} \alpha}$
is given by 
\cite{Pines}
\begin{eqnarray}
\omega_{{\bf q} \alpha} ^2  &=& 
\omega_{{\bf q} 0}^2  + \Omega_{{\rm dia}, \alpha}^2  + 
4 \pi \Pi_{\alpha \alpha} (\omega), 
\label{eq5}
\end{eqnarray}
%
%  (5)
%	
where $\Omega_{{\rm dia}, \alpha}^2$ and $4 \pi \Pi_{\alpha \alpha} (\omega)$
are, respectively, the diamagnetic and current-current contributions to the
photon self-energy,  shown in Fig.~1
(for the Q1D model under consideration,
the explicit form of $\Pi_{\alpha \alpha} (\omega)$  and  
$\Omega_{{\rm dia}, \alpha}^2$  is given in Secs. IV\,B,\,C
and  Ref.~\cite{KupcicPB2}, respectively).
	Combining the Maxwell equations with Eq.~(\ref{eq4}),
it can be shown that the transverse dielectric function 
$\varepsilon_{\alpha } (\omega)$ satisfies the relation
\begin{eqnarray}
\big[ \omega\big(\omega + {\rm i} \eta \big) 
\varepsilon_{\alpha } (\omega)
- \omega_{{\bf q}0}  ^2 \big]
\langle \langle Q_{{\bf q} \alpha} ; Q_{-{\bf q} \alpha} 
\rangle \rangle_{\omega}
 &=& \hbar,
\label{eq6}
\end{eqnarray}
%
%  (6)
%
with	
\begin{eqnarray}
\varepsilon_{\alpha } (\omega) &=& 
1 + \frac{4 \pi {\rm i}}{\omega} \sigma_{\alpha \alpha} (\omega).
\label{eq7}
\end{eqnarray}
%
%  (7)
%
	The  optical conductivity defined by Eq.~(\ref{eq7}) is
\begin{eqnarray}
\sigma_{\alpha \alpha} (\omega ) &=& 
\frac{\rm i}{4 \pi \big( \omega + {\rm i}\eta \big) }
\big[ \Omega_{{\rm dia}, \alpha}^2  + 4 \pi \Pi_{\alpha \alpha} (\omega) \big],
\label{eq8}
\end{eqnarray}
%
%  (8)
%
irrespective of whether the electromagnetic field is treated as a
classical or a quantum field.
	This is a quite general and in many respects very useful result.
	In the general case, with several valence bands 
or with several scattering channels  in a single band,
$\Omega_{{\rm dia}, \alpha}^2$ includes one or more  diamagnetic 
contributions (depending on the number of bands intersecting the Fermi level),  
and  $\Pi_{\alpha \alpha} (\omega)$ represents
all intra- and interband current-current correlation functions.

   \begin{figure}[tb]
\centerline{    
\includegraphics[height=3pc,width=15pc]{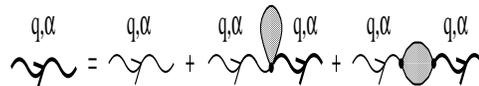} 
}   
    \caption{The Dyson equation for the photon Green function.
    The bare second-order (first-order) electron-photon coupling 
    leads to the second
    (third) term on the right-hand side, giving rise to the diamagnetic
    (current-current) contribution to the photon self-energy.    
    }
    \end{figure}

The following general properties of the expression (\ref{eq8}) are important
for interpreting  the measured spectra.
 	First, there are at least two distinct structures 
in the optical conductivity spectrum
${\rm Re} \{ \sigma_{\alpha \alpha} (\omega ) \}$,
the first one is a delta function at $\omega = 0$, related to  
the diamagnetic current, 
and the second one represents various contributions, 
including  the exciton contributions 
if the short-range dipole-dipole interactions are present
\cite{Wooten,Adler,Wieser,ZBB}.
	This is easily   seen from
\begin{eqnarray}
{\rm Re} \{  \sigma_{\alpha \alpha} (\omega ) \} &=& 
\frac{1}{4 }  \big[ 
\Omega_{{\rm dia}, \alpha}^2  + 4 \pi {\rm Re} \{ \Pi_{\alpha \alpha} (0) \} 
\big] \delta (\omega ) 
 \nonumber \\
 &&
-\frac{1}{\omega} {\rm Im } \{ \Pi_{\alpha \alpha} (\omega ) \}.
\label{eq9}
\end{eqnarray}
%
%  (9)
%
	Obviously, in the normal metallic or insulating
state the delta-function term vanishes \cite{Kittel}.
	Therefore, any consistent treatment of the electron-photon coupling 
functions has to fulfill the relation
\begin{eqnarray}
\Omega_{{\rm dia}, \alpha}^2    
 + 4 \pi {\rm Re} \{ \Pi_{\alpha \alpha} (0) \} =0 .
\label{eq10}
\end{eqnarray}
%
%  (10)
%
	The total optical conductivity in the normal metallic or insulating
state can be written  then in the form
\begin{eqnarray}
\sigma_{\alpha \alpha} (\omega ) &=& 
\frac{\rm i}{\omega }
\big[  \Pi_{\alpha \alpha} (\omega) - {\rm Re} \{ \Pi_{\alpha \alpha} (0) \} 
\big].
\label{eq11}
\end{eqnarray}
%
%  (11)
%
	Noteworthy, in the superconducting state the sum 
\begin{eqnarray}
\frac{1}{2 }  \big[ 
\Omega_{{\rm dia}, \alpha}^2  + 4 \pi {\rm Re} \{ \Pi_{\alpha \alpha} (0) \} 
\big] 
\label{eq12}
\end{eqnarray}
%
%  (12)
%
measures the  weight of the missing area in optical conductivity spectra
\cite{Schrieffer,Mattis,Tinkham}, while the single-particle contributions are
still given by Eq.~(\ref{eq11}).
	Finally, notice that in absence of local dipolar excitations
(the case of the site-energy dimerization
discussed in this article) the total current-current correlation 
function is the sum of only two contributions describing,
respectively,  the creation of the ``free'' intra- and interband 
electron-hole pairs, while the processes associated with excitons  
(i.e. the quasiparticles representing the bound electron-hole pairs) 
are not present.

	The optical conductivity determined by Eq.~(\ref{eq11}),
with the current-current correlation function  calculated  by
the equation of motion approach, is in the focus of the  present analysis.
	The causality  requirement 
\cite{Wooten,Kittel}
(i.e. the Kramers--Kronig relations),
the effective mass theorem 
\cite{KupcicPB1,Genkin,Ashcroft}
and the gauge-invariance requirement
\cite{KupcicPB1,Pines,KupcicPB2}
will be used to test the obtained results.	
	A particular care will be devoted to the non-physical singularity
at $\omega = 0$ related to the prefactor of $\omega^{-1}$ in Eq.~(\ref{eq11}),
and to the construction of the optical conductivity model with
the correct behaviour in the $\omega \rightarrow 0$ limit,
giving rise to a unified description of the optical and transport phenomena.

\section{Electronic Hamiltonian with two qualitatively different 
         scattering channels}
%
%         Sec. 3
%
Although our model is Q1D, the present response theory is quite general 
and the  electronic Hamiltonian  could represent an arbitrary 
multiband model.
	We assume  that in addition to the bare electronic Hamiltonian 
(denoted below by ${H}_0$)	
there is a static single-particle potential  
(${H}'_0$) characterized by a commensurate wave vector, which  describes, 
for example, the scattering of electrons on the site-energy dimerization 
potential. 
	The other single-electron scattering processes (on impurities,
phonons, etc.) are represented by ${H}'_1$.
	For most of the questions discussed here the main effects of the  
two-electron interactions are taken satisfactorily into account through the 
effective mean fields included in ${H}'_0$  or ${H}'_1$, \cite{footnote2}
resulting finally in $H^{\rm el}_0 = {H}_0 + {H}'_0 + {H}'_1$
in Eq.~(\ref{eq1}).
	$H^{\rm ext}=H^{\rm ext}_1 + H^{\rm ext}_2$  couples the valence
electrons to transverse electromagnetic  fields.
	The resulting total electronic Hamiltonian is
\begin{eqnarray}
H^{\rm el} &=& 
H^{\rm el}_0  + H^{\rm ext}.
\label{eq13}
\end{eqnarray}
%
%  (13)
%

We start by diagonalizing the Hamiltonian ${H}_0 + {H}'_0 $.
	As mentioned above, we consider 
the simplest Q1D  model, where ${H}'_0$  represents 
the site-energy dimerization in the highly conducting direction
and in ${H}'_1$ only the impurity scattering is taken into account.
	Next, we determine the related electron-photon coupling functions.
	At the end of this section, the multiband current-current 
correlation function is introduced.

 \subsection{Bare Hamiltonian}
%
%    Sec. 3.1
%
The single-particle properties of the Q1D site-energy-dimerization model
come from the exact diagonalization of the Hamiltonian
\cite{KupcicPB2}
\begin{eqnarray} 
{H}_0 + {H}'_0 &=&  \sum_{{\bf k} \sigma} \big[ 
\varepsilon_c ({\bf k}) c^{\dagger}_{{\bf k} \sigma} c_{{\bf k} \sigma} 
+  \varepsilon_{\underline{c}} ({\bf k}) 
\underline{c}^{\dagger}_{{\bf k} \sigma} \underline{c}_{{\bf k} \sigma}  
 \nonumber \\
 &&   
 \hspace{10mm}
+ \Delta \big(  \underline{c}^{\dagger}_{{\bf k} \sigma} c_{{\bf k} \sigma} 
+  c^{\dagger}_{{\bf k} \sigma} \underline{c}_{{\bf k} \sigma} 
\big) \big].
\label{eq14}
\end{eqnarray}
%
%  (14)
%
	The bare electron dispersions of two subbands,
artificially dimerized  along the highly conducting direction $a$, are
\begin{eqnarray}
\varepsilon_{\underline{c},c} ({\bf k})  &=&  
\pm 2t_{ a } \cos {\bf k} \cdot {\bf a} 
- 2t_{ b} \cos {\bf k} \cdot {\bf b},
\label{eq15}
\end{eqnarray}
%
%  (15)
%
with $\varepsilon_{\underline{c}} ({\bf k}) \equiv 
\varepsilon_{c} ({\bf k} \pm \pi/a \hat{x}) $,
and with the wave vector {\bf k} restricted to the new (reduced)
Brillouin zone, 
$-0.5\pi/a \le k_x \le 0.5 \pi/ a$, $-\pi/b \le k_y \le \pi/b$.
	$t_{a }$ and $t_{ b}$ ($t_{ a } \gg t_{ b} > 0$) 
are the bond energies 
in the  direction $a$ and the perpendicular direction $b$,
respectively, and $\Delta $ is the magnitude of the dimerization potential
in the direction  $a$.
	Note that such potential corresponds to imperfect nesting in
Eq.~(\ref{eq15}) in contrast to dimerization in all directions; 
the former nesting is chosen because it is more interesting 
in context of the conductivity studies.

The transformations between the unperturbed states, the band index  
$l = c, \underline{c}$, and the Bloch states, the band index  
$L = C, \underline{C}$, of the form  
\begin{eqnarray}
 l_{{\bf k} \sigma}^{\dagger} &=& 
 \sum_L U_{\bf k} (l,L)  L_{{\bf k} \sigma}^{\dagger}, 
\label{eq16}
\end{eqnarray}
%
%  (16)
% 
lead to
\begin{eqnarray}
  H_0 &=& 
  \sum_{L {\bf k} \sigma}  E_L ({\bf k}) 
  L_{{\bf k} \sigma}^{\dagger} L_{{\bf k} \sigma},
\label{eq17}
\end{eqnarray}	
%
%  (17)
%  
with the dispersions
\begin{eqnarray}
E_{\underline{C}, C} ({\bf k}) &=& 
\frac{1}{2}[\varepsilon_{\underline{c}}({\bf k}) + \varepsilon_c({\bf k})]
\pm  \sqrt{\frac{1}{4}\varepsilon_{\underline{c}c}^2  ({\bf k}) 
+ \Delta ^2}, 
\nonumber \\
\varepsilon_{\underline{c}c}  ({\bf k})  &=&  
\varepsilon_{\underline{c}} ({\bf k}) - \varepsilon_c ({\bf k}). 
\label{eq18}
\end{eqnarray}
%
%  (18)
% 
	The transformation-matrix elements are given by
\begin{eqnarray} 
 \left( \begin{array}{ll} 
  U_{\bf k} (c,C) & U_{\bf k} (c,\underline{C})  \\
  U_{\bf k} (\underline{c},C) & U_{\bf k} (\underline{c},\underline{C}) 
\end{array} \right)   
&=&  \left( \begin{array}{cc}
  u ({\bf k}) &  v ({\bf k})  
\\    -v({\bf k})  &  u({\bf k}) 
\end{array} \right),  
\label{eq19}
 \end{eqnarray} 
%
%  (19)
%  
where 
$ u ({\bf k}) =\displaystyle \cos \frac{\varphi ({\bf k})}{2}$,
$v ({\bf k}) = \displaystyle \sin \frac{\varphi ({\bf k})}{2}$,
with the auxiliary phase 
$\varphi ({\bf k})$ 
defined in the usual way
\begin{eqnarray}
\tan \varphi ({\bf k} ) &=&  \frac{2\Delta }{
\varepsilon_{\underline{c}c} ({\bf k}) }.
\label{eq20}
\end{eqnarray} 
%
%  (20)
% 
	The bands are shown in Fig.~2.
Hereafter, the lower (conduction) band is assumed to be partially filled 
and the upper (valence) band is empty.
	For the lower band completely filled and $\Delta \gg t_{b}$,
the band structure  corresponds to  the commensurate CDW system, 
otherwise we have the metallic behaviour.

    \begin{figure}[tb]
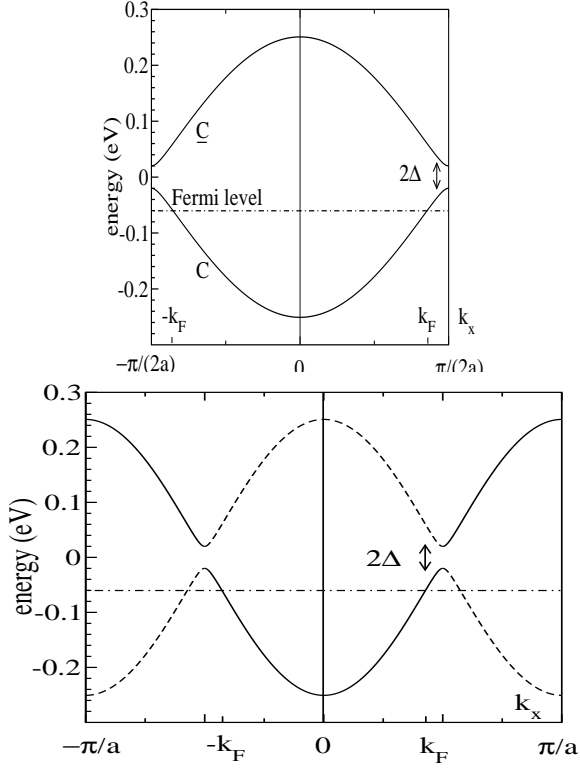

%    \centerline{
    \includegraphics[height=12pc,width=12pc]{figkupcic_2a.eps}
    \includegraphics[height=12pc,width=18pc]{figkupcic_2b.eps}    
%    }
    \caption{    
    The electron dispersions $E_{\underline{C},C} (k_x, 0.5\pi/b)$
    in the reduced (a) and extended (b) zone representation,
    for $2t_a = 0.25$ eV and $\Delta  = 0.02$ eV.
    In Fig.~(b), the solid and dashed lines correspond, respectively,  
    to $u({\bf k})$ and $v({\bf k})$ in the relations
    $C^{\dagger}_{{\bf k} \sigma} =u({\bf k}) c^{\dagger}_{{\bf k} \sigma}
    -v({\bf k}) \underline{c}^{\dagger}_{{\bf k} \sigma}$,
    $\underline{C}^{\dagger}_{{\bf k} \sigma} =v({\bf k}) 
    c^{\dagger}_{{\bf k} \sigma}
    +u({\bf k}) \underline{c}^{\dagger}_{{\bf k} \sigma}$.
    The dot-dashed line is the Fermi level $\mu$
    in a typical metallic case.
    }
    \end{figure}

 \subsection{Electron-photon coupling Hamiltonian}
%
%    Sec. 3.2
%
Using the generalized minimal substitution method for the tight-binding
electrons
\cite{KupcicPB1,KupcicPB2}, 
we obtain that the conduction electrons 
described by the  Hamiltonian (\ref{eq14}) are 
 coupled to the external electromagnetic fields through
\begin{eqnarray}
H^{\rm ext} &=& 
\sum_{l }  \sum_{{\bf k} \sigma}  \sum_{{\bf q} \alpha} 
\delta H_0^{l} ({\bf k}, {\bf q}) 
 l_{{\bf k} + {\bf q} \sigma}^{\dagger} l_{{\bf k} \sigma},
\label{eq21}
\end{eqnarray}
%
%  (21)
% 
where, to the second order in the vector potential $ A_{\alpha} ({\bf q}) $,
\begin{eqnarray}
\delta  H_0^{l} ({\bf k}, {\bf q})  &\approx &  
 - \frac{\partial  \varepsilon_{l} ({\bf k})}{\partial k_{\alpha}}
\frac{e}{\hbar c}  A_{\alpha} ({\bf q}) 
% \nonumber \\
%  &&
+ \frac{1}{2} 
\frac{\partial^2 \varepsilon_{l} ({\bf k})}{\partial k_{\alpha}^2}
\bigg( \frac{ e}{\hbar c}\bigg)^2  A_{\alpha}^2  ({\bf q}).  
\nonumber \\ \label{eq22}
\end{eqnarray}
%
%  (22)
%
	The photon annihilation operator $A_{{\bf q}\alpha}$ 
enters in Eq.~(\ref{eq22}) through
\begin{eqnarray}
&& A_{\alpha}  ({\bf q}) \hspace{2mm} =  \hspace{2mm} 
\sqrt{\frac{4 \pi c^2}{V}} Q_{{\bf q} \alpha},
 \nonumber \\
&&
A_{\alpha}^2  ({\bf q}) \hspace{2mm} = \hspace{2mm} 
\sum_{{\bf q}'} A_{\alpha}  ({\bf q}- {\bf q}') A_{\alpha}  ({\bf q}'),
\nonumber
\end{eqnarray}
where  
\begin{eqnarray}
Q_{{\bf q} \alpha} &=& 
\sqrt{\frac{\hbar}{2 \omega_{{\bf q} \alpha}}}
\big[ 
A_{{\bf q}\alpha} +  A^{\dagger}_{-{\bf q}\alpha} 
\big]
\label{eq23}
\end{eqnarray}
%
%  (23)
%
is the electromagnetic displacement field 
of Eq.~(\ref{eq2}) \cite{Pines}.
	Finally, in the Bloch representation the coupling Hamiltonian 
becomes
\begin{eqnarray}
H^{\rm ext} &=& H^{\rm ext}_1 +  H^{\rm ext}_2 
=   -\frac{1}{c} \sum_{{\bf q} \alpha}  
A_{\alpha} ({\bf q}) \hat{J}_{\alpha} (-{\bf q})
\nonumber \\
&& \hspace{5mm}
+ \frac{{ e}^2}{2mc^2} \sum_{{\bf q} \alpha} 
A^2_{\alpha} ({\bf q}) \hat{\gamma}_{\alpha \alpha} (-{\bf q};2) ,
\label{eq24}
\end{eqnarray}
%
%  (24)
%
with  
\begin{eqnarray}
\hspace{8mm}\hat{J}_{\alpha} ({\bf q})  &=&  \sum_{LL'} \sum_{{\bf k} \sigma}
J^{LL'}_{\alpha} ({\bf k})  L_{{\bf k} \sigma}^{\dagger} 
L'_{{\bf k}+{\bf q}  \sigma}, 
\label{eq25} \\ \nonumber 
\hspace{8mm}\hat{\gamma}_{\alpha \alpha} ({\bf q};2)  &=&  
\sum_{LL'} \sum_{{\bf k} \sigma}
\gamma^{LL'}_{\alpha \alpha } ({\bf k};2)  L_{{\bf k} \sigma}^{\dagger} 
L'_{{\bf k}+{\bf q}  \sigma}, 
\hspace{5mm} (25')
\end{eqnarray}
%
%  (25)
%
representing, respectively, the current density and 
Raman \cite{Genkin,KupcicPB2} density operators.
	The structure of the related vertex functions
$J^{LL'}_{\alpha} ({\bf k})$ and $\gamma^{LL'}_{\alpha \alpha } ({\bf k};2) $
is given in Appendix A.

    \begin{figure}[tb]
\centerline{
    \includegraphics[height=15pc,width=20pc]{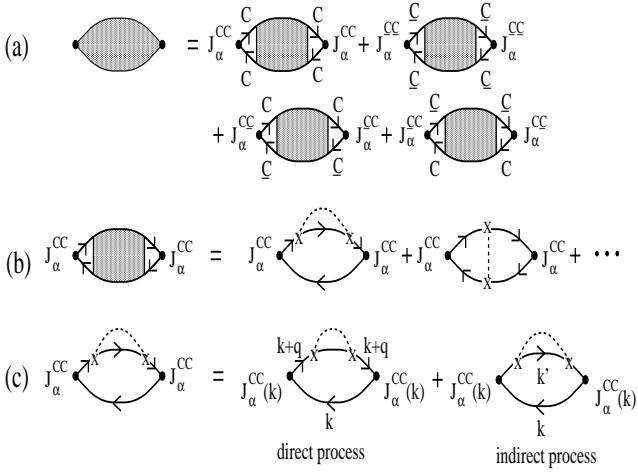} 
}      
    \caption{    
    (a) The Bloch representation of the current-current correlation function
    of the two-band model (\ref{eq14}),
    with the current vertices given by Eqs.~(\ref{eqA2}) and (A2$'$). 
    The shaded square is the electron-hole propagator defined by 
    Eq.~(\ref{eq28}).
    (b) The intraband current-current correlation function at high-frequencies
    (the leading term in powers of $(H_1')^2/\omega$).
    (c) The structure of the high-frequencies self-energy term.
    The direct  processes are associated with the creation of an electron-hole
    pair with wave vectors {\bf k} and ${\bf k}+{\bf q}$ (${\bf q}$ 
    is the wave vector of the external field), while the indirect processes  
    involve the impurity-assisted electron-hole pairs with wave vectors 
    ${\bf k}$ and ${\bf k}'+{\bf q}\approx {\bf k}'$, or 
    ${\bf k}+{\bf q}\approx {\bf k}$ and ${\bf k}'$ (${\bf k}'-{\bf k}$ 
    is the momentum relaxed on impurities, and the crosses represent 
    the impurity scattering).    
    }
    \end{figure}

\subsection{Single-particle scattering processes }
%
%   Sec. 3.3
%
We consider here only the intraband impurity scattering processes
in the perturbation  
\begin{eqnarray}
  H_1' &=& \sum_{L}
  \sum_{ {\bf k} {\bf k}'\sigma}  V^{LL} ({\bf k},{\bf k}') 
  L_{{\bf k} \sigma}^{\dagger} L_{{\bf k}' \sigma},
\label{eq26}
\end{eqnarray}
%
%  (26)
% 
for which one usually assumes 
$V^{LL} ({\bf k},{\bf k}') = V^{LL} ({\bf k}-{\bf k}')$
\cite{Abrikosov,Doniach}.
	This form of $H_1'$ is consistent with the regime in which 
$\varepsilon_{\underline{c}c} ({\bf k}_{\rm F}) 
\gg \hbar/\tau$ ($\tau$ is the relaxation 
time defined below).
	The generalization is straightforward.

\subsection{Current-current correlation function }
%
%    Sec. 3.4
%

In the equation of motion approach 
\cite{Pines} 
used in the next section, the starting point is the 
current-current correlation function of Eqs.~(\ref{eq8})--(\ref{eq12})
shown in Fig.~3.
	It is defined by 
\cite{LRA,Wooten,Mahan,KupcicPB2}
\begin{eqnarray}
\Pi_{\alpha \alpha} ({\bf q}, t)  &=& \frac{1}{\hbar V}
\langle \langle \hat{J}_{\alpha} ({\bf q}) ; \hat{J}_{\alpha} (-{\bf q})
\rangle \rangle_{t} 
\nonumber \\
 &\equiv & -\frac{{\rm i}}{\hbar V} \;
\Theta(t) \langle \big[  \hat{J}_{ \alpha}({\bf q},t), 
\hat{J}_{\alpha} (-{\bf q},0) \big] \rangle.
\label{eq27}
\end{eqnarray}
%
%  (27)
%
	Here the current operator $\hat{J}_{\alpha} ({\bf q})$ 
includes all intra- and interband current density fluctuations, as seen
from Eq.~(\ref{eq25}).
	According to Fig.~3, $\Pi_{\alpha \alpha} ({\bf q}, t)$ comprises	
two intraband and two interband contributions, 
and the problem is reduced, as will be seen immediately below, 
to the self-consistent calculation 
of  intra- and interband  electron-hole propagator in presence 
of the perturbation $H_1'$.
	In Eq.~(\ref{eq27}), as well as in 
Eqs.~(\ref{eq28}), (\ref{eq30}), (\ref{eq33}) and (\ref{eq34}),
the usual notation for the retarded correlation functions is used:
$\langle \langle \hat{A},\hat{B} \rangle \rangle_{t} =
-{\rm i} \Theta(t) \langle \big[ \hat{A}(t),\hat{B} (0)
\big] \rangle$, with $\hat{A}(t)$ being the operator $\hat{A}$ in the
Heisenberg picture.

\section{Equation of motion  approach}
%
%    Sec. 4
%

% 
%
\subsection{Generalized correlation functions}
%
%   Sec. 4.1
%

The retarded electron-hole propagator
\begin{eqnarray}
&& \! \!  \! \!  \! \!  \! \!
{\cal D}_1^{LL'}  ({\bf k},  {\bf k}_+, {\bf k}_+',{\bf k}',  t) 
=  \langle \langle 
L_{{\bf k} \sigma}^{\dagger} L'_{{\bf k}+{\bf q}  \sigma} ; 
{L'}_{{\bf k}' +{\bf q} \sigma}^{\dagger} L_{{\bf k}' \sigma}
\rangle \rangle_{t} 
\nonumber \\
&& \hspace{5mm} \equiv
 -{\rm i} 
\Theta(t) \langle \big[ 
\big( L_{{\bf k} \sigma}^{\dagger} L'_{{\bf k}+{\bf q}  \sigma} \big)_t, 
\big({L'}_{{\bf k}' +{\bf q} \sigma}^{\dagger} L_{{\bf k}' \sigma} \big)_0
] \rangle
 \nonumber \\
&& \hspace{5mm} =
{\rm e}^{ -\eta t} 
\frac{1}{2 \pi} \int_{-\infty}^{\infty} {\rm d} \omega \,
{\rm e}^{{\rm i} \omega t} 
{\cal D}_1^{LL'}  ({\bf k}, {\bf k}_+',{\bf k}_+, {\bf k}',  \omega) 
\label{eq28}
\end{eqnarray}
%
%   (28)
%
(${\bf k}_+ $ is  the abbreviation for ${\bf k} + {\bf q}$)
is the central quantity  to all long-wavelength correlation functions,
as can be seen from Fig.~4, or from the expression
\begin{eqnarray} 
&& \! \!  \! \!  \! \!  \! \!
\chi_{f, g} ({\bf q}, t) = \frac{1}{\hbar V}
\sum_{LL'} \sum_{{\bf k} {\bf k}' \sigma} 
f^{LL'} ({\bf k},{\bf k}+{\bf q}) g^{L'L} ({\bf k}' +{\bf q},{\bf k}')
\nonumber \\
&& \hspace{20mm} \times 
{\cal D}_1^{LL'}  ({\bf k},  {\bf k}_+, {\bf k}_+',{\bf k}',  t), 
\label{eq29}
\end{eqnarray}
%
%   (29)
%
which represents a generalized long-wavelength correlation function, with 
$f^{LL'} ({\bf k},{\bf k}+{\bf q})$ and  
$g^{L'L} ({\bf k}' +{\bf q},{\bf k}')$ being the charge,
current,  Raman or some other vertex functions.
	In Eq.~(\ref{eq28}) the abbreviation 
$(\hat{A}\hat{B})_t = \hat{A}(t) \hat{B}(t)$ is used.

    \begin{figure}[tb]
\centerline{
    \includegraphics[height=3pc,width=8pc]{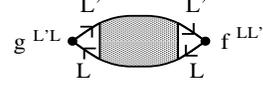}
}   
    \caption{    
    The generalized correlation function in notations of Fig.~3.
    }
    \end{figure}

The symmetry of the $f^{LL'} ({\bf k},{\bf k}+{\bf q})$ and  
$g^{L'L} ({\bf k}' +{\bf q},{\bf k}')$ vertices, together with the nature 
of the singularity of the leading term in the perturbation series, 
determines the summation rule for the related Feynman diagrams.
	The simplest way to collect the most singular diagrams 
is to consider the equations of motion connecting  
${\cal D}_1^{LL'}  ({\bf k}, {\bf k}_+, {\bf k}_+', {\bf k}', t)$ 
with the correlation function 
${\cal D}_2^{LL'}  ({\bf k},  {\bf k}_+, {\bf k}_+', {\bf k}',  t)$
defined as
\begin{eqnarray}
&& \! \!  \! \!  \! \!  \! \!
{\cal D}_2^{LL'}  ({\bf k},  {\bf k}_+, {\bf k}_+', {\bf k}',  t) 
\label{eq30}\\ \nonumber 
&& \hspace{15mm} =
\langle \langle 
\big[L_{{\bf k} \sigma}^{\dagger} L'_{{\bf k}+{\bf q}  \sigma},H'] ; 
{L'}_{{\bf k}' +{\bf q} \sigma}^{\dagger} L_{{\bf k}' \sigma}
\rangle \rangle_{t}. 
\end{eqnarray}
%
%   (30)
%
	The direct calculation gives the exact relation
\begin{eqnarray}
 && \hbar \big[ {\cal D}^{LL'}_0 ({\bf k},{\bf k}_+,\omega) \big]^{-1} 
{\cal D}_1^{LL'}  ({\bf k}, {\bf k}_+, {\bf k}_+', {\bf k}',  \omega) 
\label{eq31}
 \\ \nonumber 
 &&
 \hspace{5mm} 
= \hbar \delta_{{\bf k}, {\bf k}'} 
\big[f_L ({\bf k}) - f_{L'} ({\bf k}_+) \big]
+ {\cal D}_2^{LL'}  ({\bf k}, {\bf k}_+, {\bf k}_+', {\bf k}',  \omega).
\end{eqnarray}	
%
%   (31)
%
	Here 
\begin{eqnarray}
\hbar \big[ {\cal D}^{LL'}_0 ({\bf k},{\bf k}_+,\omega) \big]^{-1}  = 
\hbar (\omega + {\rm i} \eta) + E_L ({\bf k}) - E_{L'} ({\bf k}_+)
\nonumber \\ \label{eq32}
\end{eqnarray}
%
%   (32)
%
is a useful abbreviation.
	${\cal D}_2^{LL'}  ({\bf k}, {\bf k}_+,  {\bf k}_+',{\bf k}',  \omega)$ 
is the Fourier transform of 
${\cal D}_2^{LL'}  ({\bf k},  {\bf k}_+,  {\bf k}_+',{\bf k}',  t)$,
and $f_L ({\bf k}) \equiv f (E_L ({\bf k}))  = 
\langle {L}_{{\bf k} \sigma}^{\dagger} L_{{\bf k} \sigma} \rangle
= \big[1 +\exp \{ \beta [E_L ({\bf k} )- \mu ] \} \big]^{-1}$
is the Fermi--Dirac function, with $\beta = 1/(k_{\rm B}T)$.

The way to evaluate 
${\cal D}_2^{LL'} ({\bf k}, {\bf k}_+, {\bf k}_+',{\bf k}', t)$
depends on the choice of representation of this function.
	There are two alternative ways,
\begin{eqnarray}
&& {\cal D}_2^{LL'} ({\bf k}, {\bf k}_+, {\bf k}_+', {\bf k}', t) 
\label{eq33} 
 \\ \nonumber 
 && 
 \hspace{10mm} 
=  -{\rm i} \Theta(t) \langle \big[  \big[ 
L_{{\bf k} \sigma}^{\dagger} L'_{{\bf k}+{\bf q}  \sigma}, H' \big] _t, 
\big( {L'}_{{\bf k}' +{\bf q} \sigma}^{\dagger} L_{{\bf k}' \sigma} \big)_0] 
\rangle
\end{eqnarray}	
%
%   (33)
%
or 
\begin{eqnarray}
&&{\cal D}_2^{LL'} ({\bf k}, {\bf k}_+, {\bf k}_+', {\bf k}', t) 
\label{eq34}
  \\ \nonumber 
 && 
 \hspace{5mm} 
=  -{\rm i} \Theta(t) \langle \big[  \big[ 
L_{{\bf k} \sigma}^{\dagger} L'_{{\bf k}+{\bf q}  \sigma}, H' \big]_0, 
\big( {L'}_{{\bf k}' +{\bf q} \sigma}^{\dagger} L_{{\bf k}' \sigma}\big)_{-t}] 
\rangle,
\end{eqnarray}	
%
%   (34)
%
leading to two different self-consistent schemes (one described below 
and another encountered in the longitudinal response theory \cite{Pines}),
both giving, as will be argued below, the same result.	
	Again, $[\hat{A},\hat{B}]_t$ is the abbreviation for 
$[\hat{A}(t), \hat{B}(t)]$.

It is important to realize at the outset that the first term on the right-hand side
of Eq.~(\ref{eq31}) is significant for  all interband correlation functions
$\chi_{f, g} ({\bf q}, t)$ in Eq.~(\ref{eq29}),
irrespective of the vertex symmetries.
	The probability for the  direct creation of the interband electron-hole
pair is proportional here to 
$f_L ({\bf k}) - f_{L'} ({\bf k}_+)  \approx f_L ({\bf k}) - f_{L'} ({\bf k})$,
so that all occupied states  in the conduction band(s) are equally important.
	For  vertices $f^{LL'} ({\bf k},{\bf k}+{\bf q})$ and  
$g^{L'L} ({\bf k}' +{\bf q},{\bf k}')$ taken to represent 
the intraband current vertices, the correlation function 
$\chi_{f, g} ({\bf q}, t)$ 
in Eq.~(\ref{eq29}) becomes the intraband current-current correlation
function  $\Pi^{\rm intra} _{\alpha \alpha} ({\bf q}, t)$.
	According to the discussion of Sec.\,II\,A, the related 
intraband optical conductivity  is ruled by the prefactor  $\omega^{-1}$ 
in Eq.~(\ref{eq11}).
	The direct processes in Eq.~(\ref{eq31}) 
(see the first term on the right-hand side of Fig.~3(c))
related to  $f_C ({\bf k}) - f_{C} ({\bf k}_+) \rightarrow 0$
are insignificant, and 
$\Pi^{\rm intra} _{\alpha \alpha} ({\bf q}, t) $
can be adequately described by the second,  indirect term in 
this equation (the second term in Fig.~3(c); see also Fig.~5), 
with  ${\cal D}_2^{LL'} ({\bf k}, {\bf k}_+, {\bf k}_+', {\bf k}', t)$ 
given by Eq.~(\ref{eq34}).

   \begin{figure}[tb]
    \centerline{\includegraphics[height=13pc,width=15pc]{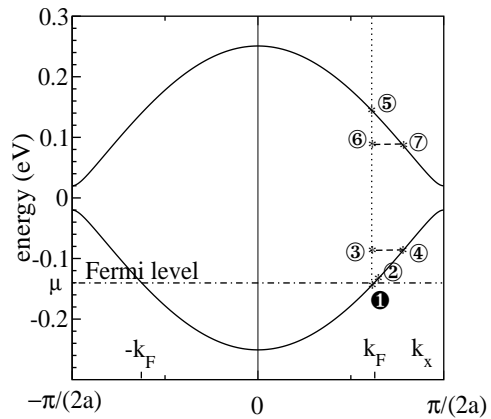}}
    \caption{    
    The  direct intraband ($1 \rightarrow 2$), 
    indirect intraband ($1 \rightarrow 3 \rightarrow 4$), 
    direct interband ($1 \rightarrow 5$) and indirect interband 
    ($1 \rightarrow 6 \rightarrow 7$)
    optical excitations. 
    In this article  the second and third processes are considered,
    the indirect intraband, using the exact summation of the impurity 
    scattering processes, and the direct interband, 
    using the phenomenological treatment of the  impurity scattering.
    The dotted line is the photon dispersion, and the dashed line corresponds
    to the impurity scattering.
    }
    \end{figure}

Before turning to the evaluation of 
$\Pi^{\rm intra} _{\alpha \alpha} ({\bf q}, t)$ 
it is interesting to contrast this conclusion with its analog in the
longitudinal response theory.
	In the longitudinal approach, the long-range Coulomb forces 
are activated, and the first term in Eq.~(\ref{eq31}) dominates the 
intraband charge-charge correlation function, even in the dynamic limit,
since the small probability for the intraband electron-hole pair creation,
proportional to $f_C ({\bf k}) - f_{C} ({\bf k}_+)$, cancels out
the $q^{-2}$ singularity of the long-range forces,
which is the well-known RPA result.
	The indirect scattering processes, on the other hand, 
are proportional to the effective intraband charge vertex, analogous 
to the effective intraband current vertex introduced below, Eq.~(\ref{eq41}).
	Since the bare long-wavelength intraband charge vertex satisfies
$q({\bf k}+{\bf q},{\bf k}) \approx q({\bf k},{\bf k}) = e$,
where $e$ is the bare electron charge, the
effective intraband charge vertex for the indirect scattering processes 
vanishes, because
$q({\bf k},{\bf k}) - q({\bf k}',{\bf k}') = 0$. 
	Since the longitudinal and transverse approaches are to be equivalent,  
this means  that the direct longitudinal processes are to be equivalent 
to the indirect transverse processes in the intraband channel, 
while the contributions of both the indirect longitudinal 
and the direct transverse scattering processes are to be negligible.
	This issue will be further discussed in Sec.\,IV\,B\,5.

The above transverse approach can now be applied to the current-current 
correlation function of the two band model, rewritten in the form
\begin{eqnarray}
\Pi_{\alpha \alpha} ({\bf q}, t)  &=& \frac{1}{\hbar V}
\sum_{LL'} \sum_{{\bf k} {\bf k}' \sigma} 
J^{LL'}_{\alpha} ({\bf k}) J^{L'L}_{\alpha} ({\bf k}')
 \nonumber \\
 && \hspace{5mm} 
 \times
{\cal D}_1^{LL'}  ({\bf k},  {\bf k}_+, {\bf k}_+',{\bf k}', t). 
\label{eq35}
\end{eqnarray}
%
%   (35)
% 
Here the vertices $f^{LL'} ({\bf k},{\bf k}+{\bf q})$  and 
$g^{L'L} ({\bf k}' +{\bf q},{\bf k}')$ in Eq.~(\ref{eq29}) are
replaced by $J^{LL'}_{\alpha} ({\bf k},{\bf k}+{\bf q})\approx 
J^{LL'}_{\alpha} ({\bf k})$  and 
$J^{L'L}_{\alpha} ({\bf k}' +{\bf q},{\bf k}') \approx 
J^{L'L}_{\alpha} ({\bf k}')$, respectively.
	Optical processes relevant to the two-band model 
(including the indirect interband ones not considered in the present 
analysis) are illustrated in Fig.~5.

In order to make presentation of the results more transparent,
we shall first determine the intraband contributions, 
and then give the  analysis of the interband optical excitations.
	For the sake of brevity, in the next section the 
(intra)band  index $C$ will be omitted 
($E_{C} ({\bf k}) \rightarrow E ({\bf k})$,
$J^{CC}_{\alpha} ({\bf k}) \rightarrow J_{\alpha} ({\bf k})$, \ldots,
with $C_{{\bf k} \sigma}^{\dagger} \rightarrow c_{{\bf k} \sigma}^{\dagger}$).
	It should be noticed that the results obtained below for the 
intraband optical conductivity are quite general, i.e. they 
cover various physically different  regimes.
	As explained  in Sec.\,IV\,B\,2, when the electron 
filling of the dimerized band  varies between 0 and 1, 
the electronic system transforms from an electron-like semiconducting, 
through a metallic, into a hole-like semiconducting regime.
	A more detailed discussion of this issue is given in Sec.\,V,
in  context of the total optical conductivity.

\subsection{Intraband optical conductivity}
%
%    Sec. 4.2
%

According to the aforementioned arguments, the intraband optical processes
are described by the equations
\begin{eqnarray}
&& \! \!  \! \!  \! \!  \! \!
\hbar {\cal D}_0^{-1}( {\bf k},{\bf k}_+,\omega) 
{\cal D}_1({\bf k}, {\bf k}_+, {\bf k}_+', {\bf k}',  \omega) 
\nonumber \\
&& \hspace{25mm} = 
{\cal D}_2 ({\bf k}, {\bf k}_+,  {\bf k}_+',{\bf k}',  \omega),
\label{eq36}
 \\
&&  \! \!  \! \!  \! \!  \! \!
\hbar  {\cal D}_0^{-1} ({\bf k}',{\bf k}'_+,\omega) 
{\cal D}_2 ({\bf k}, {\bf k}_+, {\bf k}_+', {\bf k}',  \omega) 
\nonumber \\
&& \hspace{10mm} =
- {\cal D}_3({\bf k}, {\bf k}_+, {\bf k}_+', {\bf k}',  \omega)
\nonumber \\
&& \hspace{15mm}
\label{eq37} 
+ \hbar \langle \big[  \big[ 
c_{{\bf k} \sigma}^{\dagger} c_{{\bf k}+{\bf q}  \sigma}, H' \big], 
{c}_{{\bf k}' +{\bf q} \sigma}^{\dagger} c_{{\bf k}' \sigma}] 
\rangle,
\end{eqnarray}
%
%   (36-37)
%
where ${\cal D}_0^{-1} ({\bf k}',{\bf k}'_+,\omega)$ is the intraband term 
in Eq.~(\ref{eq32})  and
${\cal D}_3({\bf k}, {\bf k}_+,  {\bf k}_+',{\bf k}',  \omega)$ 
is the Fourier transform of the force-force correlation function 
\cite{Mahan,Gotze,KupcicPB2}
\begin{eqnarray}
&& 
{\cal D}_3({\bf k},  {\bf k}_+,  {\bf k}_+',{\bf k}',  t) 
\label{eq38}
 \\ \nonumber 
 && 
 \hspace{5mm} 
=
  -{\rm i} \Theta(t) \langle 
\big[  \big[ 
c_{{\bf k} \sigma}^{\dagger} c_{{\bf k}+{\bf q}  \sigma}, H' \big]_0, 
\big[ 
c_{{\bf k}' +{\bf q} \sigma}^{\dagger} c_{{\bf k}' \sigma},\, H' \big]_{-t}] 
\rangle.
\end{eqnarray}
%
%   (38)
%
	The second term in Eq.~(\ref{eq37}) is the ground-state  average of the 
four-operator product at $t=0$. 
	It is off-diagonal in the Bloch representation, and its value 
can be obtained by putting $E ({\bf k}'_+) -  E ({\bf k}') \approx 0$
in the left-hand side  of Eq.~(\ref{eq37})
and then taking the formal limit $\omega \rightarrow 0$.
	The result is 
\begin{eqnarray}
&&  \! \!  \! \!  \! \!  \! \!
\hbar \langle \big[  \big[ 
c_{{\bf k} \sigma}^{\dagger} c_{{\bf k}+{\bf q}  \sigma}, H' \big],  
{c}_{{\bf k}' +{\bf q} \sigma}^{\dagger} c_{{\bf k}' \sigma}] 
\rangle
\nonumber \\
&& \hspace{20mm} 
\approx 
{\cal D}_3({\bf k}, {\bf k}_+, {\bf k}_+', {\bf k}', \omega  = 0).
\label{eq39}
\end{eqnarray}
%
%   (39)
%
 	For the impurity scattering, we have the  cancellation of   
this constant term with its counterpart obtained by the electron
$\rightleftharpoons$ hole replacement (see Eq.~(\ref{eq47}), for example), 
but for a time-dependent perturbation $H_1'$, 
Eq.~(\ref{eq39}) achieves an interesting structure, as pointed out in Refs.  
\cite{Gotze,Giamarchi}

Due to the large velocity of light, 
the energy and wave vector transfers in the external points of the
intraband current-current correlation function  fulfill 
$\hbar \omega  \gg  E ({\bf k}'_+) -  E ({\bf k}')$, and the factors 
${\cal D}_0 ({\bf k},{\bf k}_+, \omega )$ and 
${\cal D}_0({\bf k}',{\bf k}'_+, \omega)$
in Eqs.~(\ref{eq36}) and (\ref{eq37}), representing the propagator of the 
virtual electron-hole pairs (related to the process $1 \rightarrow 3$ in
Fig.~5), can be replaced by  $1/\omega$.
	The resulting intraband  correlation function becomes
\begin{eqnarray}
&&\Pi^{\rm intra}_{\alpha \alpha} (\omega)  = 
- \frac{1}{\hbar V(\hbar \omega)^2}
\sum_{{\bf k} {\bf k}'' {\bf k}_1 {\bf k}''_1 \sigma} 
\big[ J_{\alpha} ({\bf k}) - J_{\alpha} ({\bf k}'') \big]  
\nonumber \\
&&\hspace{10mm}\times 
\big[ J_{\alpha} ({\bf k}''_1) - J_{\alpha} ({\bf k}_1) \big]  
\bigg<  V({\bf k} - {\bf k}'') V({\bf k}''_1 - {\bf k}_1)  
\nonumber  \\ 
&& \times 
\big[ 
{\cal D}_1({\bf k}, {\bf k}'', {\bf k}_1'', {\bf k}_1,  \omega) 
- {\cal D}_1({\bf k}, {\bf k}'', {\bf k}_1'', {\bf k}_1, 0) \big] \bigg> .
\label{eq40} 
\end{eqnarray}
%
%   (40)
%
	$\langle \ldots \rangle$ is here and subsequently 
the average over the impurity sites \cite{Mahan}.

    \begin{figure}[tb]
\centerline{
    \includegraphics[height=10pc,width=20pc]{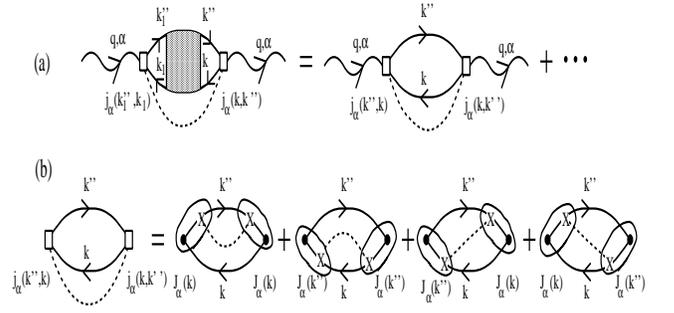}
}
    \caption{    
    (a) The indirect contributions to the
    intraband current-current correlation function in the equation 
    of motion approach.    
    The open square is the effective current vertex, and
    the dashed line represents the related force-force correlation functions
    (see, for example, the discussion of the force-force correlation function
    method in  Sec. 8.1.B of Ref.~\cite{Mahan}).
    (b) 
    The leading term in the high-frequency limit.
    On the right-hand side, the products 
    $V({\bf k} - {\bf k}'')J_{\alpha} ({\bf k}) /(\hbar \omega)$ and
     $V({\bf k} -{\bf k}'')J_{\alpha} ({\bf k}'') /(\hbar \omega)$
    are encircled into the effective vertex.
    The dot is the bare intraband current vertex. 
    }
    \end{figure}

The diagrammatic representation of
$\Pi^{\rm intra}_{\alpha \alpha} ({\bf q}, \omega)$ is given in Fig.~6(a),
with the structure of the leading term, proportional to 
$(H_1')^2/\omega$, shown explicitly in Fig.~6(b).
	The expression 
\begin{eqnarray}
j_{\alpha} ({\bf k}, {\bf k}'') &=&  
\frac{V({\bf k} - {\bf k}'')}{\hbar \omega}  \big[ J_{\alpha} ({\bf k}) 
- J_{\alpha} ({\bf k}'') \big] 
\label{eq41}
\end{eqnarray}
%
%   (41)
%
can be recognized as an  effective current vertex for the indirect
intraband photon absorption/emission processes,
and is a sum of two terms encircled in Fig.~6(b). 
	Notice also that 
${\cal D}_1({\bf k}, {\bf k}'',  {\bf k}_1'',{\bf k}_1,  \omega) 
- {\cal D}_1({\bf k}, {\bf k}'',  {\bf k}_1'', {\bf k}_1, 0)$ 
is proportional to $\omega$, canceling out the factor of $\omega^{-1}$ 
in one of the effective vertices.

It is important to notice that 
in the leading term (Fig.~6(b))
the expression (\ref{eq40}) is identical to the result of  
the force-force correlation function method  \cite{Mahan}.
	While the latter method is usually limited to the examination 
of this leading term, 
or to the summation of irrelevant higher order diagrams 
which results in the well-known Hopfield formula \cite{Mahan},
the present approach is focused instead on the exact summation 
of the most singular contributions  in powers of $(H_1')^2/\omega$ 
and should be regarded as a generalization
of the standard force-force correlation function approach.

\subsubsection{Proper electron-hole representation}
%
%   Sec. 4.2.1
%

After determining the structure of the external points (i.e.~the effective
vertices) in the diagram shown in Fig.~6(a), 
we have to find the internal structure of the diagram.
	The latter is  represented by 
the indirect (impurity-assisted) electron-hole propagator 
${\cal D}_1  ({\bf k}, {\bf k}'', {\bf k}_1'', {\bf k}_1,  \omega)$,
characterized by the momentum transfer ${\bf k} - {\bf k}''$, 
rather than by the negligibly small external momentum transfer 
${\bf k}_+ - {\bf k} = {\bf q}$ of the direct processes
in Eq.~(\ref{eq31}).
	Its internal structure is determined here by the self-consistent 
solution of the exact equation 
\begin{eqnarray}
&& \! \!  \! \!  \! \!  \! \! 
\hbar {\cal D}^{-1}_0  ({\bf k},{\bf k}'', \omega) 
{\cal D}_1  ({\bf k}, {\bf k}'', {\bf k}_1'', {\bf k}_1,  \omega) 
\label{eq42}
\\  \nonumber
&&  
= \hbar \delta_{{\bf k}, {\bf k}_1} \delta_{{\bf k}'', {\bf k}_1''} 
\big[f ({\bf k}) - f ({\bf k}'') \big] 
+ {\cal D}_2 ({\bf k}, {\bf k}'', {\bf k}_1'', {\bf k}_1,  \omega).
\end{eqnarray}
%
%   (42)
%
	${\cal D}_1  ({\bf k}, {\bf k}'', {\bf k}_1'', {\bf k}_1,  \omega) $
represents  the electron-hole pair created by the indirect photon
absorption of Fig.~5, which means that 
the energy transfer $\hbar \omega$ is 
close to the electron-hole pair energy $ E({\bf k}) -  E({\bf k}'')$.
	The wave vectors {\bf k} and ${\bf k}''$ 
(${\bf k}_1$ and ${\bf k}''_1$, as well) are independent of each other, and  
therefore the first term   in Eq.~(\ref{eq42}) 
dominates the low-energy physics, preferring the optical processes
between the states $E({\bf k}) \approx \mu$ and $E({\bf k}'') \approx  \mu$.

    \begin{figure}[tb]
\centerline{    
    \includegraphics[height=13pc,width=20pc]{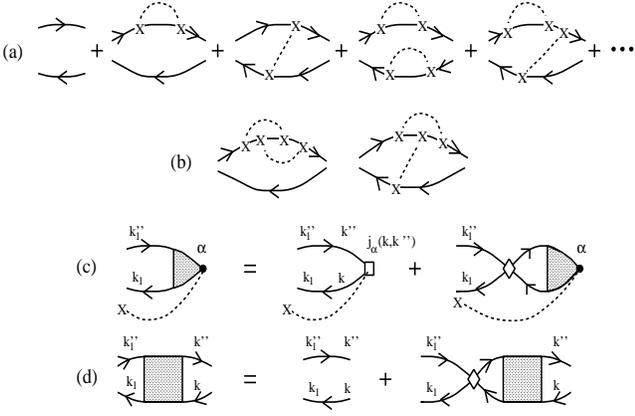}
}    
    \caption{ (a) A few leading contributions included in the self-consistent
    treatment of Eq.~(\ref{eq42}).
    (b) Typical contributions omitted in the self-consistent scheme.    
    (c) The self-consistent equation of the kernel 
    ${\cal F}_{\alpha}  ({\bf k}''_1, {\bf k}_1,  \omega)$  and
    (d)  of the electron-hole propagator 
    ${\cal D}_1  ({\bf k}, {\bf k}'', {\bf k}_1'', {\bf k}_1,  \omega) $.
    The diamond represents the electron-hole self-energy
    consisting of the single-particle self-energy and vertex-correction
    contributions.}
    \end{figure}

In  combining Eq.~(\ref{eq42}) with the equation of motion for
${\cal D}_2 ({\bf k}, {\bf k}'', {\bf k}_1'', {\bf k}_1,  \omega)$
(see Eq.~(\ref{eq53}) for more detail)
it is sufficient to collect only the terms in the 
summation which are relevant to the self-consistent  equation for
the kernel ${\cal F}_{\alpha} ({\bf k}_1'', {\bf k}_1,  \omega)$
in the current-current correlation function (see Figs.~7(c,d),
and the criterion of the validity shown in Figs.~7(a,b))
\begin{eqnarray}
&&\Pi^{\rm intra}_{\alpha \alpha} (\omega) =
- \frac{1}{\hbar V (\hbar \omega)^2} 
\sum_{{\bf k}_1 {\bf k}''_1 \sigma} 
\bigg< \big[ J_{\alpha} ({\bf k}''_1) - J_{\alpha} ({\bf k}_1) \big] 
\nonumber \\ 
&& 
\hspace{15mm}  \times
\big[ 
{\cal F}_{\alpha}({\bf k}_1'', {\bf k}_1,  \omega) 
- {\cal F}_{\alpha}({\bf k}_1'', {\bf k}_1, 0) \big]\bigg>,
\label{eq43} 
\end{eqnarray}
%
%   (43)
%
where
\begin{eqnarray}
&& \! \!  \! \!  \! \!  \! \!
{\cal F} _{\alpha} ({\bf k}_1'', {\bf k}_1,  \omega)
= 
\sum_{{\bf k} {\bf k}''} 
V({\bf k} - {\bf k}'')   V({\bf k}''_1 - {\bf k}_1)
\label{eq44} \nonumber \\  
 && \hspace{5mm} \times 
\big[ J_{\alpha} ({\bf k}) - J_{\alpha} ({\bf k}'') \big]
{\cal D}_1  ({\bf k}, {\bf k}'', {\bf k}_1'', {\bf k}_1, \omega).
\end{eqnarray}
%
%   (44)
%
	The idea of the present approach is the self-consistent treatment
of Eqs.~(\ref{eq42}) and (\ref{eq44}), which represent two equations 
connecting the electron-hole propagator 
${\cal D}_1  ({\bf k}, {\bf k}'', {\bf k}_1'', {\bf k}_1, \omega)$
with the kernel 
${\cal F} _{\alpha} ({\bf k}_1'', {\bf k}_1,  \omega)$. 
	The solution is based on the expansion
\begin{eqnarray}
{\cal D}_1  ({\bf k}, {\bf k}'', {\bf k}_1'', {\bf k}_1,  \omega) = 
\sum_{n=0}^{\infty} 
{\cal D}_1^{(2n)}  ({\bf k}, {\bf k}'', {\bf k}_1'', {\bf k}_1,  \omega),  
\label{eq45}
\end{eqnarray}
%
%   (45)
%
where ${\cal D}_1^{(2n)}  ({\bf k}, {\bf k}'', {\bf k}_1'', {\bf k}_1,  \omega)$
includes only the self-consistent terms proportional to $(H_1')^{2n}/\omega^n$.

As easily seen in the longitudinal analysis, 
the self-consistent expression for 
${\cal D}_1  ({\bf k}, {\bf k}'',  {\bf k}_1'', {\bf k}_1,  \omega)$,
with ${\bf k}'' \rightarrow {\bf k}+{\bf q}$, plays the leading role.
	Accordingly, to compare both response theories, one needs the 
self-consistent scheme for  
${\cal F} _{\alpha} ({\bf k}_1'', {\bf k}_1,  \omega)$ and
the recurrence relations for 
${\cal D}_1^{(2n)}  ({\bf k}, {\bf k}'', {\bf k}_1'', {\bf k}_1,  \omega)$
(see Sec.\,IV\,B\,3).

	For pedagogical reasons, it is convenient first to consider the 
zeroth order contribution  to (\ref{eq43})
and  define the effective number of conduction
electrons and the related electron-hole damping energy.

\subsubsection{High-frequency limit}
%
%  Sec. 4.2.2
%

The direct calculation gives the first term in the expansion (\ref{eq45})
\begin{eqnarray} 
&& {\cal D}_1^{(0)}  ({\bf k}, {\bf k}'', {\bf k}_1'', {\bf k}_1,  \omega) 
\label{eq46}
 \\ \nonumber 
 && \hspace{10mm} 
= 
\delta_{{\bf k}, {\bf k}_1}  \delta_{{\bf k}'', {\bf k}_1''} 
\big[f ({\bf k}) - f ({\bf k}'') \big] 
{\cal D}_0  ({\bf k},{\bf k}'', \omega).
\end{eqnarray}
%
%  (46)
%
	The related contribution to  the current-current correlation function
(which is a good approximation for 
$\Pi^{\rm intra}_{\alpha \alpha} (\omega)$ 
at high frequencies) is given by
\begin{eqnarray}
&& \Pi^{\rm intra, (0)}_{\alpha \alpha} (\omega)  = 
\frac{1}{\big( \hbar \omega \big)^2}  
\frac{1}{V} \sum_{{\bf k} {\bf k}'' \sigma} 
J_{\alpha} ({\bf k})
\big(  J_{\alpha} ({\bf k}) - J_{\alpha} ({\bf k}'') \big)
\nonumber \\
&& \hspace{10mm} \times 
\big[ f ({\bf k}) - f ( {\bf k}'') \big]
\langle | V({\bf k} - {\bf k}'') | ^2 \rangle 
 \nonumber \\
 && \hspace{20mm}\times 
 \frac{1}{\hbar} \big[  {\cal D}_0 ({\bf k}, {\bf k}'', \omega) 
- {\cal D}_0 ({\bf k}'', {\bf k}, \omega) \big] 
\label{eq47}
\\
 && \hspace{10mm} \approx
\frac{1}{ \big(\hbar \omega \big)^2}  
\frac{1}{V} \sum_{{\bf k} {\bf k}'' \sigma} 
J_{\alpha} ({\bf k})
\big(  J_{\alpha} ({\bf k}) - J_{\alpha} ({\bf k}'') \big) 
\nonumber \\
&& \hspace{15mm}\times 
\big[ f (E({\bf k})) - f ( E({\bf k}) + \hbar \omega) \big] 
\langle | V({\bf k} - {\bf k}'') | ^2 \rangle
 \nonumber \\
 && \hspace{20mm}\times 
 \frac{1}{\hbar}  \big[  {\cal D}_0 ({\bf k}, {\bf k}'', \omega) 
+ {\cal D}_0 ({\bf k}'', {\bf k}, \omega) \big] . 
\hspace{5mm}(47')
\nonumber 
\end{eqnarray}
%
%  (47)
% 
	For the impurity scattering processes, the real part of this function 
is negligible, 	while  the imaginary part is given by
\begin{eqnarray}
 &&{\rm Im} \{ \Pi^{\rm intra (0) }_{\alpha \alpha} (\omega) \}  
\approx 
- \frac{ 1 }{\big( \hbar \omega \big)^2}  \frac{1}{V}
 \sum_{{\bf k}  \sigma} J^2_{\alpha} ({\bf k}) 
\label{eq48}
 \\ \nonumber 
 && \hspace{20mm} \times  
\big[ f (E({\bf k})) - f ( E({\bf k}) + \hbar \omega) \big] 
\frac{\hbar}{\tau ({\bf k}, \omega)},
\nonumber
\end{eqnarray}
%
%  (48)
%
where  the electron-hole damping energy is
\begin{eqnarray}
 && \frac{\hbar}{\tau ({\bf k},  \omega)}   
 = 
 \sum_{{\bf k}'' } \langle | V({\bf k} - {\bf k}'') | ^2 \rangle
\bigg(  
1 - \frac{J_{\alpha} ({\bf k}'') }{ J_{\alpha} ({\bf k})}
\bigg) 
\nonumber \\
&& \hspace{20mm} \times
 (-) \frac{2}{\hbar} {\rm Im} 
 \big\{  {\cal D}_0 ({\bf k}, {\bf k}'', \omega) \big\}  .
\label{eq49}
\end{eqnarray}
%
%  (49)
% 
	Furthermore, in this case, the frequency dependent part in 
$1 / \tau ({\bf k},\omega)$ is negligibly small, and we can put		
the average over the Fermi surface 
$\langle 1 / \tau ({\bf k},0) \rangle_{\rm FS} \equiv 1/\tau  $ 
in Eq.~(\ref{eq48}) instead of 
$1 / \tau ({\bf k},\omega)$. 
	We finally get
\cite{KupcicPB2}
\begin{eqnarray}
&&{\rm Im} \{ \Pi^{\rm intra, (0) }_{\alpha \alpha} ( \omega) \}  
\approx 
-\frac{e^2  n^{\rm eff}_{\rm intra, \alpha} }{m} 
\frac{1  }{\omega \tau },
\label{eq50}
\end{eqnarray} 
%
%  (50)
%
with
\begin{eqnarray}
n^{\rm eff}_{\rm intra, \alpha}  &=& 
-\frac{ m }{ e^2}  \frac{1}{V}
\sum_{{\bf k}  \sigma} J^2_{\alpha} ({\bf k}) 
\frac{\partial  f ({\bf k})}{\partial  E({\bf k})} 
\label{eq51}
\end{eqnarray} 
%
%  (51)
%
being the effective number of conduction electrons.
	The behaviour of $n^{\rm eff}_{\rm intra, a}$ with band filling 
$\delta$ is  shown in Fig.~8 for a few typical values of the ratio 
$\Delta/(2t_{a})$, and compared to the prediction of the free-electron and
free-hole approximation.
	Notice that for $\Delta > 2t_{a} \gg 2t_{b}$ one obtains the 
well-known result 
$n^{\rm eff}_{\rm intra, a} \propto \sin k_{{\rm F}x}2a= \sin \delta \pi$
($2a$ is the primitive-cell parameter of the dimerized lattice).

    \begin{figure}[tb]
\centerline{
    \includegraphics[height=14pc,width=18pc]{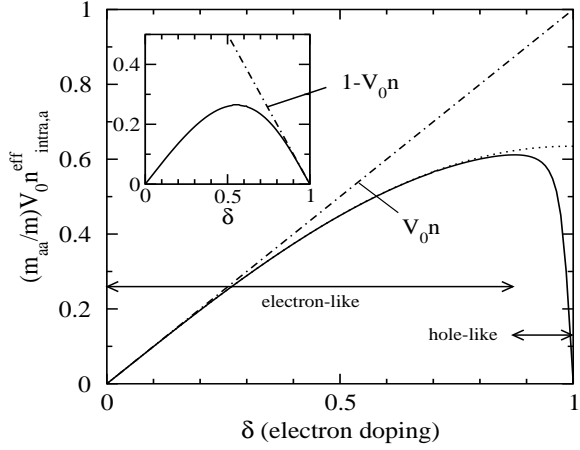}    
}
    \caption{The effective number of conduction electrons (\ref{eq51})
    shown in the dimensionless form, for      
    $2t_a = 0.25$ eV, $t_b \rightarrow 0$, $T = 10$ K and 
    $\Delta = 0$ (dotted line), 10 meV (solid line).
    The dot-dashed line is the dimensionless concentration 
    of conduction electrons 
    $V_0 n =  2a k_{\rm F} /\pi \equiv \delta$, i.e.
    the prediction of the free-electron approximation. 
    $V_0$ is the primitive cell volume and $m_{aa} = \hbar^2/(2t_a a^2)$ 
    is the mass scale parameter.
    The doping regions with the electron-like 
    ($\partial n_{{\rm intra},a}^{\rm eff}/ \partial \delta > 0$)
    and hole-like
    ($\partial n_{{\rm intra},a}^{\rm eff}/ \partial \delta < 0$)
    conductivity are also indicated.
    Inset: The effective number $(m_{aa}/m)V_0 n_{{\rm intra},a}^{\rm eff}$
    for $2t_a = 0.25$ eV, $t_b \rightarrow 0$, $T = 10$ K and 
    $\Delta = 0.25$ eV.
    The dot-dashed line is the prediction of the free-hole approximation,
    $1-V_0n$.
    }
    \end{figure}

Beyond this approximation the expression (\ref{eq47}) 
requires the evaluation of two coupled integrations over {\bf k} and 
${\bf k}''$.
	In the low-dimensional electronic systems, where the van Hove
singularities in the band structure may play important role,
this is not a trivial task.

The most outstanding advantage of the present  approach is
the fact that the effective current vertex 
$j_{\alpha} ({\bf k}, {\bf k}'')$ consists of two terms, which, together with two
terms in $j_{\alpha} ({\bf k}_1'', {\bf k}_1)$, give four contributions to
$\Pi^{\rm intra, (0)}_{\alpha \alpha} (\omega)$,
two of them representing the so-called self-energy contributions, and 
the other two the vertex corrections (see Fig.~6(b))
\cite{Mahan,Gotze}.
	We shall show next that in the $n$th order  term in (\ref{eq45}) 
the single-particle vertex corrections and the single-particle 
self-energy contributions are also treated on equal footing,
even if the relaxation processes on impurities are treated beyond the
approximation $1 / \tau ({\bf k},\omega) \approx 1/\tau $.

\subsubsection{Kernel in the low-frequency limit}
%
%   4.2.3
%

The effects of the impurity scattering on the correlation function 
(\ref{eq37}) can be exhibited in the following way:
\begin{eqnarray}
&& \! \! \! \! \! \! \! \! 
{\cal D}_2 ({\bf k}, {\bf k}'', {\bf k}_1'', {\bf k}_1,  \omega) 
= \sum_{\bf q}  V({\bf q} ) 
\big[
{\cal D}_1  ({\bf k}, {\bf k}''- {\bf q}, {\bf k}_1'', {\bf k}_1,  \omega)
 \nonumber \\
 && 
 \hspace{25mm}  
- {\cal D}_1  ({\bf k}+ {\bf q}, {\bf k}'', {\bf k}_1'', {\bf k}_1,  \omega) 
\big].
\label{eq52} 
\end{eqnarray}
%
%  (52)
%	
	When combined with this expression, Eqs. (\ref{eq42}) and 
(\ref{eq46}) lead to 
\begin{eqnarray}
&&\! \! \! \! \! \! \! \! 
\label{eq53}
{\cal D}_1  ({\bf k}, {\bf k}'', {\bf k}_1'', {\bf k}_1,  \omega) 
\approx
{\cal D}_1^{(0)}  ({\bf k}, {\bf k}'', {\bf k}_1'', {\bf k}_1,  \omega)
\\
&& \hspace{20mm}
+
\frac{1}{\hbar} {\cal D}_0  ({\bf k},{\bf k}'', \omega)
 \sum_{\bf q}   | V({\bf q} ) | ^2 
\nonumber \\
&& \hspace{10mm} \times
\frac{1}{\hbar} \big\{
{\cal D}_0  ({\bf k},{\bf k}''+{\bf q}, \omega)+
{\cal D}_0  ({\bf k}+{\bf q},{\bf k}'', \omega) 
\big\}
\nonumber \\
&&  \times
\big[
{\cal D}_1  ({\bf k}, {\bf k}'', {\bf k}_1'', {\bf k}_1,  \omega)
-
{\cal D}_1  ({\bf k}+{\bf q}, {\bf k}''+{\bf q}, {\bf k}_1'', {\bf k}_1,  \omega)
\big].
\nonumber 
\end{eqnarray}
%
%   (53)
%
	On the right-hand side of this equation, only  the self-consistent 
terms are taken into account.
	The first term in the brackets represents the single-particle
self-energy contributions and the second one the single-particle
vertex corrections.
	It is important to remember again that these corrections are the
largest for $E({\bf k}) \approx E({\bf k}'')$, so that 
${\cal D}_0  ({\bf k},{\bf k}'', \omega) $ 
can be approximated by $\omega^{-1}$.
	As a consequence, the solution of this equation can be sought 
in powers of $ | V({\bf q} ) | ^2 /  \omega$.
	However, due to the dependence on 
${\bf k}+{\bf q}$ and $ {\bf k}''+{\bf q}$ of the vertex-corrections 
term, we have to turn back to the kernel (\ref{eq44}) and apply
several changes to its  dummy variables to obtain the self-consistent description
of ${\cal F} _{\alpha} ({\bf k}_1'', {\bf k}_1,  \omega)$ and 
the desired recurrence relations between the contributions 
${\cal D}_1 ^{(2n)} ({\bf k}, {\bf k}'', {\bf k}_1'', {\bf k}_1,  \omega)$.
	The kernel is described by 
\begin{eqnarray}
&&{\cal F} _{\alpha} ({\bf k}_1'', {\bf k}_1,  \omega)
- {\cal F} _{\alpha}^{(0)}  ({\bf k}_1'', {\bf k}_1,  \omega)
\nonumber  \\
 && \hspace{10mm} =
\frac{1}{\omega}\sum_{{\bf k} {\bf k}''} 
\big| V({\bf k} - {\bf k}'') \big|^2 
\big[ J_{\alpha} ({\bf k}'') - J_{\alpha} ({\bf k}) \big]
\nonumber \\  
&& \hspace{15mm} \times 
{\cal D}_1  ({\bf k}, {\bf k}'', {\bf k}_1'', {\bf k}_1,  \omega)
\Sigma({\bf k}, {\bf k}'', \omega),
\label{eq54} 
\end{eqnarray}
%
%   (54)
%
where
\begin{eqnarray}
&& \hbar \Sigma({\bf k}, {\bf k}'', \omega) 
\label{eq55} \\ \nonumber 
&& \hspace{5mm} 
= 
-\sum_{{\bf q} } \langle | V({\bf q} - {\bf k}'') | ^2 \rangle
\bigg(  
1 - \frac{J_{\alpha} ({\bf q}) }{ J_{\alpha} ({\bf k}'')}
\bigg) 
\frac{1}{\hbar} {\cal D}_0 ({\bf k}, {\bf q}, \omega) 
\nonumber \\  
&& \hspace{10mm}  
- \sum_{{\bf q} } \langle | V({\bf q} - {\bf k}) | ^2 \rangle
\bigg(  
1 - \frac{J_{\alpha} ({\bf q}) }{ J_{\alpha} ({\bf k})}
\bigg) 
\frac{1}{\hbar} {\cal D}_0 ({\bf q}, {\bf k}'', \omega)  
\nonumber 
\end{eqnarray}
%
%   (55)
%
is the electron-hole self-energy and 
$ {\cal F} _{\alpha}^{(0)}  ({\bf k}_1'', {\bf k}_1,  \omega)$
is the kernel corresponding to the replacement of 
${\cal D}_1  ({\bf k}, {\bf k}'', {\bf k}_1'', {\bf k}_1,  \omega)$
in Eq.~(\ref{eq44}) by
${\cal D}_1^{(0)}  ({\bf k}, {\bf k}'', {\bf k}_1'', {\bf k}_1, \omega)$.
	The only approximation made in the derivation of Eq. (\ref{eq54})  
is
\begin{eqnarray}
&&  J_{\alpha} ({\bf k}-{\bf k}''+ {\bf q}) -  J_{\alpha} ({\bf q})
% \\ \nonumber 
% && 
% \hspace{20mm} 
\approx
\big[ 
J_{\alpha} ({\bf k}) -  J_{\alpha} ({\bf k}'')
\big] 
\frac{J_{\alpha} ({\bf q}) }{ J_{\alpha} ({\bf k}'')},
\nonumber \\
\label{eq56}
\end{eqnarray}
%
%   (56)
%
which allows a simple description of the vertex corrections in 
${\cal F} _{\alpha} ({\bf k}_1'', {\bf k}_1,  \omega)$,
and which treats correctly the disappearance of the forward scattering
contributions (${\bf k} \approx {\bf k}''$) to both 
$\Sigma({\bf k}, {\bf k}'', \omega) $ and 
${\cal F} _{\alpha} ({\bf k}_1'', {\bf k}_1,  \omega)$.
	As briefly discussed at the end of Sec.\,IV\,B\,5,
this approximation is directly related to the restrictions 
enforced by the  the continuity equation for the charge density.
	Similarly, the definition (\ref{eq45}), together with the
self-consistent relations (\ref{eq53}) and (\ref{eq54}), 
gives rise to the recurrence relations illustrated in Fig.~7(d):
\begin{eqnarray}
&&{\cal D}_1^{(2n)}  ({\bf k}, {\bf k}'', {\bf k}_1'', {\bf k}_1,  \omega) 
\label{eq57}\\
&& \hspace{15mm} =
\frac{-\Sigma({\bf k}, {\bf k}'', \omega) }{ \omega}
{\cal D}_1^{(2n-2)}  ({\bf k}, {\bf k}'', {\bf k}_1'', {\bf k}_1,  \omega)
\nonumber \\
&&  \hspace{15mm}=  
\bigg(
\frac{-\Sigma({\bf k}, {\bf k}'', \omega) }{ \omega}
\bigg)^{n}
{\cal D}_1^{(0)}  ({\bf k}, {\bf k}'', {\bf k}_1'', {\bf k}_1,  \omega).
\nonumber 
\end{eqnarray}
%
%   (57)
%   

% 
%
\subsubsection{Memory-function approximation}
%
%   Sec. 4.2.4
%
The simplest way to solve Eq.~(\ref{eq54}) is to replace 
$\Sigma({\bf k}, {\bf k}'', \omega)$ by its imaginary part averaged over the 
Fermi surface, ${\rm i} /\tau (\omega) $.	
	As mentioned above, for the impurity scattering processes, the 
real part of $\Sigma({\bf k}, {\bf k}'', \omega)$ can be ignored.
	Even if ${\rm Re} \{\Sigma({\bf k}, {\bf k}'', \omega)\}$  
is not small, we can turn back to the  electronic Hamiltonian
and tray to include the ${\rm Re} \{\Sigma({\bf k}, {\bf k}'', \omega)\}$
effects into the  effective single-particle Hamiltonian, and to diagonalize 
it, as we did here with the scattering processes on the
dimerization potential $H_0'$.
	The real part of the new self-energy 
$\Sigma({\bf k}, {\bf k}'', \omega)$ is minimized in this way, with only 
the imaginary part playing an important role in Eq.~(\ref{eq57}).

In this case, we obtain the expression
\begin{eqnarray}
 {\cal F} _{\alpha} ({\bf k}_1'', {\bf k}_1,  \omega) &\approx&
{\cal F} _{\alpha}^{(0)}  ({\bf k}_1'', {\bf k}_1,  \omega)
\frac{\omega }{
\omega +{\rm i}  /\tau  (\omega)},
\label{eq58}
\end{eqnarray}
%
%   (58)
%
which leads to 	the well-known results of  the memory-function 
approximation
\cite{KupcicPB2,Gotze,Giamarchi}
\begin{eqnarray}
\Pi^{\rm intra  }_{\alpha \alpha} (\omega)  
&\approx& 
-\frac{{ e}^2  n^{\rm eff}_{\rm intra, \alpha} }{m} 
\frac{{\rm i}   /\tau  (\omega) }{
\omega + {\rm i}   /\tau   (\omega)  },
\nonumber \\
\sigma^{\rm intra }_{\alpha \alpha} (\omega)  
&\approx& 
 \frac{\rm i}{\omega} \frac{{ e}^2  n^{\rm eff}_{\rm intra, \alpha} }{m} 
\frac{\omega }{\omega + {\rm i}   /\tau   (\omega)  },
\label{eq59}
\end{eqnarray}
%
%   (59)
%
with $\hbar/\tau  (\omega)$ being the intraband memory (relaxation) function.
 	This result is consistent with the causality requirement,
\begin{eqnarray}
{\rm Re} \{ \sigma^{\rm intra }_{\alpha \alpha} (-\omega) \} &=&
{\rm Re} \{ \sigma^{\rm intra }_{\alpha \alpha} (\omega) \}, 
\nonumber \\
{\rm Im} \{ \sigma^{\rm intra }_{\alpha \alpha} (-\omega) \} &=&
-{\rm Im} \{ \sigma^{\rm intra }_{\alpha \alpha} (\omega) \} ,
\label{eq60}
\end{eqnarray}
%
%   (60)
%
provided that $\tau (\omega) = \tau$.
	If $\tau (\omega)$ is frequency dependent, 
the corresponding ${\rm Re} \{\Sigma({\bf k}, {\bf k}'', \omega)\}$ 
is  non-zero, but the result (\ref{eq59}) is still  acceptable.
	For ${\rm Re} \{\Sigma({\bf k}, {\bf k}'', \omega)\}$
not too large we can introduce the effects of 
${\rm Re} \{\Sigma({\bf k}, {\bf k}'', \omega)\}$ through
the  mass redefinition $m \rightarrow m(\omega)$
through the Kramers--Kronig relations.
	The result is the  generalized Drude formula for the intraband
optical conductivity.

Here we show two important results.
	First, the memory-function approximation, which in the traditional 
form has not been found to be transparent,
can be understood as a simple replacement
$\Sigma ({\bf k},\omega ) \rightarrow {\rm i} /\tau (\omega)$ of 
the exact-summation result given below.
	Second, the memory-function results are acceptable even in the
cases with the pronounced optical excitations across a gap (or pseudogap) (where 
$ n^{\rm eff}_{\rm intra, \alpha}  \ll n$; see, for example, 
the hole-like semiconducting regime in Fig.~8 at $\delta \approx 1$), 
provided that the electron group
velocity $v_{\alpha} ({\bf k}) = J_{\alpha} ({\bf k})/e$ in 
Eq.~(\ref{eq51}) is determined using the relation (\ref{eqA5}).
	The intraband conductivity spectrum obtained in this way 
is related with the interband  conductivity spectrum 
through the well-controlled conductivity 
sum rule \cite{KupcicPB2}.

\subsubsection{Exact summation}
%
%   Sec. 4.2.5
%
In the case where  the wave vector dependence 
of the imaginary part of $\Sigma({\bf k}, {\bf k}'', \omega)$ is significant 
and the real part of $\Sigma({\bf k}, {\bf k}'', \omega)$ 
is not too large
the  full recurrence relations for 
${\cal D}_1^{(2n)}  ({\bf k}, {\bf k}'', {\bf k}_1'', {\bf k}_1,  \omega)$
can be used to obtain the intraband 
current-current correlation function.
	The result is
\begin{eqnarray} 
&&\Pi^{\rm intra}_{\alpha \alpha} (\omega)
= 
\frac{1}{V}
 \sum_{{\bf k} {\bf k}'' \sigma} 
\big[ J_{\alpha} ({\bf k}) - J_{\alpha} ({\bf k}'') \big]^2
\frac{ \langle | V({\bf k} - {\bf k}'') | ^2 \rangle }{
 \hbar \omega }  
\nonumber \\
&&  \hspace{15mm} \times  
\big[f ({\bf k}) - f ({\bf k}'') \big] 
\frac{1}{\hbar} \frac{ {\cal D}_0  ({\bf k},{\bf k}'', \omega)}{
  \hbar\omega +  \hbar \Sigma({\bf k}, {\bf k}'', \omega) },
\label{eq61}
 \\
&& \hspace{10mm}\approx 
\frac{1}{V} \sum_{{\bf k} {\bf k}''\sigma} 
J^2_{\alpha} ({\bf k})
\frac{\partial  f ({\bf k})}{\partial  E({\bf k})}  
 \frac{  1 }{
  \hbar\omega +  \hbar \Sigma({\bf k}, {\bf k}'', \omega) }
\nonumber \\
&& \hspace{20mm} \times   
\langle | V({\bf k} - {\bf k}'') | ^2 \rangle 
\bigg(  
1 - \frac{J_{\alpha} ({\bf k}'') }{ J_{\alpha} ({\bf k})}
\bigg)
 \nonumber \\
 && \hspace{20mm}\times 
\frac{1}{\hbar}  \big[  {\cal D}_0 ({\bf k}, {\bf k}'', \omega) 
+ {\cal D}_0 ({\bf k}'', {\bf k}, \omega) \big]. 
\hspace{5mm}(61')
\nonumber 
\end{eqnarray}
%
%  (61)
%
Physically the most important case corresponds to the approximation  
$\Sigma({\bf k}, {\bf k}'', \omega) \approx \Sigma({\bf k}, {\bf k}, \omega)
= \Sigma({\bf k}, \omega)$.
	In this case, we obtain 
\begin{eqnarray} 
&& 
\Pi^{\rm intra}_{\alpha \alpha} (\omega)%
 \approx  
-\frac{1}{V} \sum_{{\bf k} \sigma} 
J^2_{\alpha} ({\bf k})
\frac{\partial  f ({\bf k})}{\partial  E({\bf k})}  
 \frac{   \Sigma({\bf k},  \omega) }{
  \omega +   \Sigma({\bf k}, \omega) }.
\hspace{5mm}(61'')
\nonumber 
\end{eqnarray}
%
%  (61'')
%	
	The final form of  the optical conductivity comes from 
Eqs.~($61''$) and (11)
\begin{eqnarray}
&&\sigma^{\rm intra}_{\alpha \alpha} (\omega)  \approx   \frac{\rm i}{\omega} 
\frac{1}{V}
 \sum_{{\bf k}  \sigma} 
 J^2_{\alpha} ({\bf k}) 
(-) \frac{\partial f ({\bf k})}{\partial E ({\bf k})} 
\frac{ \omega }{
 \omega +  \Sigma({\bf k}, \omega) }, 
\nonumber \\
\label{eq62}
\end{eqnarray}
%
%  (62)
%
which is the result  identical to the result of the longitudinal
response theory.

The longitudinal response theory
gives a simpler way to obtain the same
$\sigma^{\rm intra}_{\alpha \alpha} (\omega)$.
	The difference between the two approaches is in the way how the 
continuity equation connecting the charge density and current density
fluctuations is treated.
	The consideration of the direct processes of the wave vector
{\bf q} in the longitudinal approach allows a more precise treatment of the 
continuity equation, in the way analogous to the Landau response 
theory \cite{Pines,KupcicUP}.
	But here only an approximate solution is possible, since
the theory is formulated in terms of 
the indirect intraband processes (of the wave vector ${\bf k}-{\bf k}''$).

\subsubsection{Zero-frequency limit}
%
%  Sec. 4.2.6
%
First significant consequence of the present exact summation method is the
behaviour of the intraband optical conductivity   
in the zero-frequency limit.
	When the real part of $\Sigma({\bf k},  \omega)$ is
small enough, we can write
\begin{eqnarray}
\Sigma({\bf k},  \omega) &\approx &
{\rm i} \Sigma''({\bf k},  0) \equiv 
{\rm i} / \tau ({\bf k}),
\label{eq63}
\end{eqnarray}
%
%  (63)
%
resulting in the DC conductivity which is equal to the well-known Boltzmann
result
\cite{Mahan,Ziman}
\begin{eqnarray}
 \sigma^{\rm intra}_{\alpha \alpha} (0) \equiv 
\sigma^{\rm DC}_{\alpha \alpha} &=&
(-)\frac{1}{V}
 \sum_{{\bf k} \sigma} 
 J_{\alpha} ^2 ({\bf k}) 
\frac{  \partial f ({\bf k})}{\partial E ({\bf k})}
\tau ({\bf k}) 
\nonumber \\
&=& 
\frac{e^2 \tau_0}{m V_0}  \frac{m}{m_{aa}}
 \tilde{n}^{\rm eff}_{ \rm intra, \alpha}.
\label{eq64}
\end{eqnarray}
%
%  (64)
%
	Here
\begin{eqnarray}
\tilde{n}^{\rm eff}_{\rm intra, \alpha}  &=&  
\frac{m_{aa}}{m}V_0 n^{\rm eff}_{\rm intra, \alpha} 
\nonumber \\
 &=&   
(-)  \frac{ m_{aa}}{N}
\sum_{{\bf k}  \sigma}  v^2_{\alpha} ({\bf k})
\frac{\partial  f ({\bf k})}{\partial  E({\bf k})} 
\frac{\tau ({\bf k})}{\tau_0},  
\nonumber \hspace{13mm} (51')
\end{eqnarray} 
%
%  (51')
%
is the effective number of conduction electrons shown in the dimensionless
form
and $\tau_0 = \tau({\bf k}=0)$ is the  temperature dependent 
${\bf k}=0$  relaxation time.
 	The relation (\ref{eq62}), together with Eq.~(\ref{eq66}),
gives the complete description of the optical conductivity in a general
multiband model, with the symmetry of the intra- and interband current vertices 
playing an essential role. 
	Thus, Eq.~(\ref{eq64}) provides the direct link
between the low-frequency optical conductivity  and various transport 
coefficients not only in the single-band but also in the multiband 
models.

\subsection{Interband optical conductivity}
%
%   Sec. 4.3
%

The approximation in which
the ${\cal D}_2  ({\bf k}, {\bf k}'', {\bf k}_1,  \omega)$
term in Eq.~(\ref{eq31}) is omitted leads to the ideal interband 
current-current 
correlation function and to the ideal  interband conductivity 
characterized by a sharp threshold at the energy
$E_{\underline{C}} ({\bf k}_{\rm F}) - E_{C} ({\bf k}_{\rm F})$
\cite{LRA,Wooten,Gruner,KupcicPB2}.
	The former one is given by
\begin{eqnarray}
\Pi_{\alpha \alpha}^{\rm inter} (\omega) &=& 
\frac{1}{V} \sum_{{\bf k} \sigma} 
|J_{{\alpha}}^{\underline{C}C} ( {\bf k})|^2  
\bigg\{  \frac{  f_C({\bf k}) - f_{\underline{C}}({\bf k})  
}{\hbar  \omega  -  E_{\underline{C}C} (  {\bf k})    + 
 {\rm i}  \hbar \eta }
\nonumber \\ 
&& 
\hspace{10mm}+
\frac{  f_{\underline{C}}({\bf k}) - f_C({\bf k}) 
}{\hbar  \omega + E_{\underline{C}C} ( {\bf k})  + 
{\rm i} \hbar \eta } 
\bigg\},
 \label{eq65} 
 \end{eqnarray}
%
%   (65)
% 
with $E_{\underline{C}C} ( {\bf k}) = 
E_{\underline{C}} ( {\bf k})- E_{C} ( {\bf k})$.
	The latter one comes from Eqs.~(\ref{eq11}) and (\ref{eq65})
\begin{eqnarray}
\sigma_{\alpha \alpha}^{\rm inter} (\omega) 
  &=&
\frac{\mathrm{i}}{\omega}
\frac{1}{V} \sum_{{\bf k} \sigma} \frac{\hbar \omega
|J_{{\alpha}}^{\underline{C}C} ( {\bf k})|^2  
}{E_{\underline{C}C}  (  {\bf k}) }
\nonumber \\
&& \hspace{5mm} \times
 \frac{ 2\big[ f_C({\bf k}) - f_{\underline{C}}({\bf k}) \big] 
}{\hbar \omega + {\rm i} \hbar \eta 
- E^2_{\underline{C}C} (  {\bf k})/\big( \hbar \omega \big) } ,
\label{eq66}
 \end{eqnarray}
%
%   (66)
% 
with $\eta = 0$ put in the numerator and $\eta \rightarrow 0^+$ in the
denominator.
	The leading term in the interband electron-hole self-energy
\begin{eqnarray}
\hbar \Sigma_{LL'} ({\bf k}, \omega) 
&=& 
-\sum_{{\bf q} } \bigg[
\langle | V^{L'L'}({\bf q} - {\bf k}) | ^2 \rangle
\frac{1}{\hbar} {\cal D}_0^{LL'} ({\bf k}, {\bf q}, \omega) 
 \nonumber \\ 
 && 
+ \langle | V^{LL}({\bf q} - {\bf k}) | ^2 \rangle
\frac{1}{\hbar} {\cal D}_0^{LL'} ({\bf q},{\bf k}, \omega)
\bigg]
\label{eq67} 
\end{eqnarray}
%
%   (67)
% 
comes from the single-particle self-energy contributions.
	A reasonable generalization for the   interband optical conductivity 
is given by Eq.~(\ref{eq66}) with  the replacement 
$ \eta$ by ${\rm Im} \{\Sigma_{\underline{C}C} ({\bf k}, \omega) \}$.

     \begin{figure}[tb]
\centerline{
    \includegraphics[height=14pc,width=18pc]{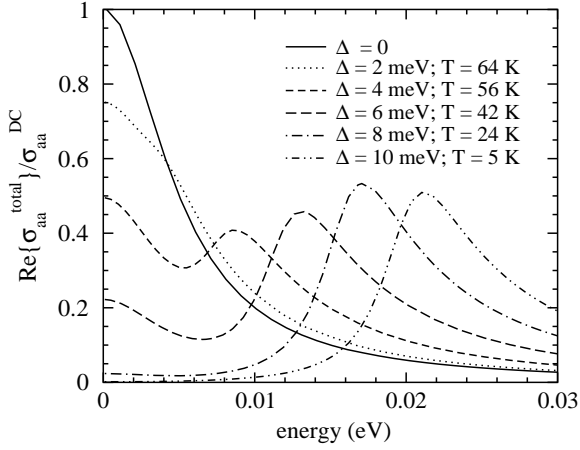}    
}
    \caption{The development of the normalized total single-particle optical 
    conductivity with temperature in the ordered CDW state: 
    $\Delta (T) = \Delta_0 \sqrt{ 1 - T/T_{\rm CDW} }$, $\Delta_0 = 10$ meV,
    $T_{\rm CDW} = 66$ K, $k_{\rm F} = 0.5 \pi/a$,      
    $2t_a = 0.25$ eV, $t_b \rightarrow 0$ and
    $\hbar \Gamma_{\rm intra} = \hbar \Gamma_{\rm inter} =$ 5 meV. 
    $\sigma_{aa}^{\rm DC} $ labels 
    the DC conductivity at $\Delta = 0$. 
    }
    \end{figure}

The  interplay between the self-energy and vertex-corrections 
terms in $\Sigma_{\underline{C}C} ({\bf k}, \omega)$, 
the correct treatment  of the indirect interband optical excitations,
as well as the role of the effective mass theorem in resolving all 
these issues  will be  explained elsewhere \cite{KupcicUP}.
	It should  be noticed here that the  Landau-like function 
(\ref{eq66}) gives the in-gap optical conductivity slightly different 
from the corresponding Lindhard-like function, as easily seen 
by comparing Fig.~9  with Fig.~4 reported in Ref.~\cite{KupcicPB2}. 
	The understanding of the difference between these two 
results is  of general importance as well, 
and will  be discussed in detail in Ref.~\cite{KupcicUP}.

    \begin{figure}[tb]
\centerline{
    \includegraphics[height=14pc,width=18pc]{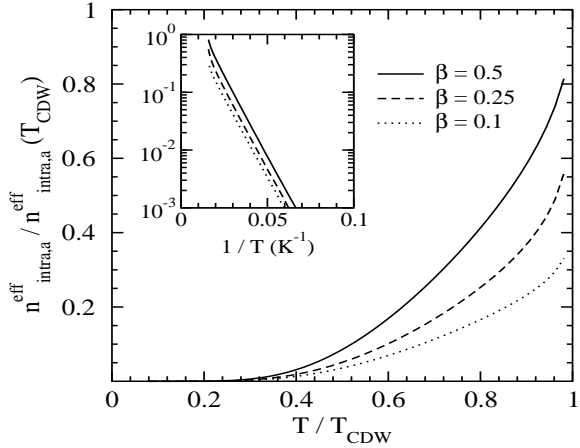}    
}
    \caption{The  temperature dependence of the 
    normalized effective number of conduction electrons 
    in the ordered CDW state for different values of exponent $\beta$: 
     $\Delta_0 = 10$ meV, $T_{\rm CDW} = 66$ K,
    $k_{\rm F} = 0.5 \pi/a$,      
    $2t_a = 0.25$ eV, $t_b \rightarrow 0$ and
    $\hbar \Gamma_{\rm intra} = \hbar \Gamma_{\rm inter} =$ 5 meV.  
    Notice that this figure also represents the temperature behaviour of the
    normalized DC conductivity, since 
    $ \sigma_{aa}^{\rm DC} (T) / 
    \sigma_{aa}^{\rm DC} (T_{\rm CDW})\approx 
    \tilde{n}_{{\rm intra}, a}^{\rm eff} (T)/
    \tilde{n}_{{\rm intra}, a}^{\rm eff} (T_{\rm CDW})$.
    }
    \end{figure}

\section{Comparison with experiments}
%
%  Sec. 5
%
In the simples limit, 
$\Sigma_{CC} ({\bf k}, \omega) \approx {\rm i} \Gamma_{\rm intra}$
and 
$\Sigma_{C\underline{C}} ({\bf k}, \omega) \approx {\rm i} 
\Gamma_{\rm inter}$, the optical conductivity of the present two-band model,
\begin{eqnarray}
 \sigma_{\alpha \alpha} (\omega)
 &=&
 \sigma^{\rm intra}_{\alpha \alpha} (\omega)  
+
 \sigma^{\rm inter}_{\alpha \alpha} (\omega),
\label{eq69}
\end{eqnarray}
%
%  (69)
%
is a function of the Fermi wave vector ${\bf k}_{\rm F}$, 
three band parameters, $t_a$, $t_b$ and $\Delta$, and two 
damping energies, $\hbar \Gamma_{\rm intra}$ and $\hbar \Gamma_{\rm inter}$.
	For $k_{{\rm F}x} \approx 0.5 \pi/a$, the model illustrates
optical properties of various Q1D imperfectly nested CDW systems 
(including both the ordered CDW state and the pseudogap effects 
at temperatures above the critical temperature $T_{\rm CDW}$).
	In this section, we shall briefly discuss a few qualitative results
important to the Q1D CDW systems.
	First, the temperature dependence of the DC and optical 
conductivity in the ordered CDW state is discussed for the strictly 1D 
case ($t_b \rightarrow 0$).
	Then we contrast the interband conductivity found here to the 
oversimplified semiconducting optical conductivity usually encountered 
in the textbooks \cite{LRA,Wooten,Gruner}.

\subsection{DC conductivity in the ordered CDW state}
%
%    Sec. 5.1
%

The temperature dependence in $k_{\rm B}T$, $\Delta (T)$, 
$\Gamma_{\rm intra} (T)$ and $\Gamma_{\rm inter} (T)$
is responsible for the  transfer with increasing/decreasing temperature
of  the optical conductivity spectra 
between the intraband and interband channels. 
	This effect  is  particularly large  in the vicinity 
of the metal-to-insulator phase transition.
	It should be noticed that, in the approximation in which 
$\tau ({\bf k}) \rightarrow  \tau_0 = 1/\Gamma_{\rm intra}$
is set in Eq.~($51'$),  the  temperature dependence of 
$\sigma_{\alpha \alpha}^{\rm DC}$
is given  by the product of  the effective number of  conduction electrons
$ \tilde{n}^{\rm eff}_{{\rm intra}, \alpha} (T)$ and the relaxation time 
$\tau_0 (T)$.
	We have two adjustable parameters at any temperature,
and the  analysis of the DC conductivity data is thus possible only 
by the combination with the optical conductivity measurements.
	The latter method allows also the determination of the 
magnitude of CDW order parameter  $\Delta_0$ and  the critical exponent $\beta$
in $\Delta (T) = \Delta_0 \big( 1 - T/T_{\rm CDW} \big)^\beta$,
as well as   the bond energy $t_a$ and the damping energy 
$\Gamma_{\rm inter}$. 

In  Figs.~9 and 10 the optical conductivity  normalized to the 
DC conductivity at $\Delta = 0$ and the temperature dependence 
of the DC conductivity are shown for  typical values of the parameters.

\subsection{Optical conductivity of a simple semiconductor}
%
%  Sec. 5.2
%

	The  typical result for the interband conductivity in the 
ordered CDW state is shown in Fig.~11 (solid and dotted curves),
and compared to the data measured in the blue bronze
K$_{0.3}$MoO$_3$ (diamonds) \cite{Degiorgi}.
	Notice that the gauge-invariance factor 
$ \hbar \omega/E_{\underline{C}C}  (  {\bf k}) $ 
in Eq.~(\ref{eq66}) ensures the disappearance of 
$\sigma_{\alpha \alpha}^{\rm inter} (0)$ at $T \rightarrow 0$,
independently of the value of the phenomenologically introduced damping energy
$\hbar \Gamma_{\rm inter}$.
	The dashed curve is the prediction of the usual optical model 
for semiconductors \cite{LRA,Wooten,Gruner}, with the factor 
$\hbar \omega/E_{\underline{C}C}  (  {\bf k}) $  absent,
which is characterized by
a significant (but non-physical) contribution to 
$\sigma^{\rm DC}_{\alpha \alpha}$ at $T \rightarrow 0$.

    \begin{figure}[tb]
\centerline{
    \includegraphics[height=14pc,width=18pc]{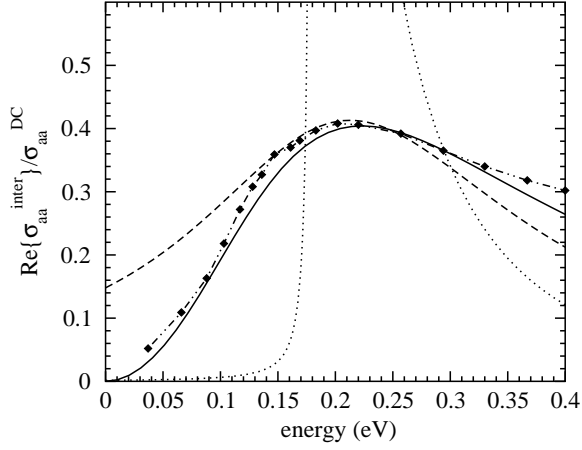}    
}
    \caption{The interband optical conductivity (\ref{eq66})    
    ($\hbar \eta \rightarrow \hbar \Gamma_{\rm inter}$),   for 
    $2t_a = 0.4$ eV, $t_b \rightarrow 0$, $\Delta = 0.09$ eV, 
    $k_{\rm F} = 0.5 \pi/a$, $T = 10$ K
    and
    $\hbar \Gamma_{\rm inter} =$ 0.3
    eV (solid curve) and 1 meV (dotted curve).
    The dashed curve is the prediction of the usual (gauge-non-invariant) 
    optical conductivity for semiconductors, calculated using the same 
    band parameters, with $\hbar \Gamma_{\rm inter} =$ 0.15 eV.
    $\sigma^{\rm DC}_{aa}$ is the DC conductivity of the 
    $\Delta = 0$,  $\hbar \Gamma_{\rm intra}  =$ 0.15 eV  case.
    The diamonds are experimental data measured in K$_{0.3}$MoO$_3$
    at $T = 10$ K \cite{Degiorgi}.
    }
    \end{figure}

\section{Conclusion}
%
%  Sec. 6
%
In this article, the response of the conduction electrons in a Q1D multiband
model to the transverse electromagnetic fields has been studied
in presence of two scattering mechanisms.
	In contrast to the coherent scattering on
the site-energy-dimerization potential, 
the impurity scattering processes give negligibly
small contribution to the real part of the electron self-energy,
but dominate in the relaxation processes in both the intra- and interband 
optical excitations.
	We determine the current-current correlation function, 
i.e. the optical conductivity
of the related two-band model as a function of band filling.
	The transverse equation of motion approach has been used to collect 
the most singular contributions to the optical conductivity.
	It is shown that the present multiband optical analysis 
represents a generalization of the usual force-force correlation 
function method, and that in the DC limit it approaches  correctly 
the results of  Boltzmann equations,
due to its gauge-invariant form. 
	The present  optical conductivity  model gives the frequency 
and temperature dependences of the single-particle contributions to
the optical conductivity spectra in the ordered CDW state which are 
consistent with both the experimental observation and the prediction of the
longitudinal response theory.
	It is  explained that, while the response to the longitudinal
fields  is associated with the direct electron-hole pair excitations, the
response to the transverse electromagnetic fields can be understood in terms 
of the indirect (impurity-assisted) electron-hole pair excitations.

\section*{Acknowledgements} 

This research was supported by the Croatian Ministry of Science and Technology 
under  the project 0119-256.
	One of the authors (I.K.) would like to acknowledge the hospitality
of the Institute of Physics of Complex Matter, Lausanne, 
where parts of this work were completed.

\vspace{5mm}

\appendix

\section{Current and static Raman vertices}
%
%  Sec. 
%
	
The  vertex functions in the expression (\ref{eq25})  depend on 
the unperturbed vertices 
$J^{ll}_{\alpha} ({\bf k}) = (e/\hbar )
\partial  \varepsilon_{l} ({\bf k}) / \partial k_{\alpha}$
and 
$\gamma^{ll}_{\alpha \alpha } ({\bf k};2) = (m/\hbar^2)
\partial^2 \varepsilon_{l} ({\bf k})/\partial k_{\alpha}^2$
in the  way
\begin{eqnarray}
J^{LL'}_{\alpha} ({\bf k}) &=& 
\sum_{l} U_{\bf k} (l,L) U^* _{\bf k} (l,L') J^{ll}_{\alpha} ({\bf k}), 
\nonumber \\
\gamma^{LL'}_{\alpha \alpha} ({\bf k};2)  &=& 
\sum_{l} U_{\bf k} (l,L) U^* _{\bf k} (l,L') 
\gamma^{ll}_{\alpha \alpha } ({\bf k};2).
\label{eqA1}
\end{eqnarray}	
%
%  (A1)
%
	For the $\alpha = a$ polarization 
of the electromagnetic fields, the result is
\begin{eqnarray}
J^{\underline{C} \underline{C}, CC}_{a} ({\bf k}) &=& 
\mp \big[ u^2 ({\bf k}) - v^2 ({\bf k}) \big] J^{cc}_{a} ({\bf k}) 
 \nonumber \\  &=& 
\mp \cos \varphi ({\bf k}) J^{cc}_{a} ({\bf k}),
\label{eqA2} \\ 
J^{\underline{C} C}_{a} ({\bf k}) &=& 2u ({\bf k}) v ({\bf k})  J^{cc}_{a} ({\bf k}) 
 \nonumber \\ &=&  
\sin \varphi ({\bf k}) J^{cc}_{a} ({\bf k}),
\hspace{30mm} ({\rm A2}')
\nonumber  
\\
\gamma^{\underline{C} \underline{C}, CC}_{a a} ({\bf k};2)  &=& 
\mp \big[ u^2 ({\bf k}) - v^2 ({\bf k}) \big] \gamma^{cc}_{a a} ({\bf k};2) 
 \nonumber \\ &=& 
\mp \cos \varphi ({\bf k}) \gamma^{cc}_{aa} ({\bf k};2).
\label{eqA3}
\end{eqnarray}	
%
%  (A2-3)
%
	Similarly, for $\alpha = b$ 
\begin{eqnarray}
J^{\underline{C} \underline{C}, CC}_{b} ({\bf k}) 
\hspace{2mm} = \hspace{2mm}J^{cc}_{b} ({\bf k}),
%\nonumber \\
&& J^{\underline{C} C}_{b} ({\bf k}) \hspace{2mm}= \hspace{2mm}0,
\nonumber  \\
\gamma^{\underline{C} \underline{C}, CC}_{bb} ({\bf k};2)  &=& 
\gamma^{cc}_{bb} ({\bf k};2).
\label{eqA4}
\end{eqnarray}	
%
%  (A4)
%
	Finally, using Eqs.~(\ref{eq18}), (\ref{eqA2}) and (\ref{eqA4})
we can check  the Ward identity 
\cite{Mahan}  
which connects the intraband current vertex $J^{LL}_{\alpha} ({\bf k})$  
with the electron group velocity $v^{L}_{\alpha} ({\bf k})$
\begin{eqnarray}
J^{LL}_{\alpha} ({\bf k})  &=& e v^{L}_{\alpha} ({\bf k}) 
\hspace{2mm}=\hspace{2mm}
\frac{e}{\hbar} \frac{\partial E_L ({\bf k})} {\partial  k_{\alpha} }.
\label{eqA5}
\end{eqnarray}	
%
%  (A5)
%


\begin{references}
\bibitem[14]{footnote1}
Notice the redundant adiabatic factor 
$\displaystyle {\rm e}^{- \eta t} $
in the Fourier transformation, which is necessary in the 
differential treatment of the retarded correlation functions.
\bibitem[24]{footnote2}
In contrast to that, in the longitudinal approach, based on  
the charge-charge correlation  functions,  
the long-range Coulomb interactions  play the crucial  role.
\end{references}
\end{document}